\documentclass[preprint]{aastex}
\usepackage{color}
\usepackage{lscape}
\usepackage{longtable,lscape}
\usepackage{rotating}
\usepackage{caption}
\usepackage{epsfig}
\usepackage{subfig}
\usepackage{graphicx}

\begin{document}

\captionsetup[subfigure]{position=bottom,width=0.86\textwidth,font=small,labelformat=empty,labelsep=none}

\newcommand{\dg}{$^{\circ}$} 
\newcommand{\longdash}{---} 
\newcommand{\hyphensym}{--} 
\newcommand{\minussign}{$-$} 
\newcommand{\sex}{{\it SExtractor}}
\newcommand{\malkrel}{MGT98 relationship}
\newcommand{\malkcit}{MGT98}
\newcommand{\ginm}{Gini-M$_{20}$}
\newcommand{\lacos}{{\tt l.a.\!cosmic}}
\newcommand{\Msol}{M$_{\odot}$}
\newcommand{\gsim}{$\gtrsim$}
\newcommand{\lsim}{$\lesssim$}
\newcommand{\txw}{\textwidth}
\newcommand{\syall}{Sy1 and Sy2}
\newcommand{\snr}{signal-to-noise ratio}
\newcommand{\hypz}{{\tt HyperZ}}
\newcommand{\gfit}{{\tt GalFit}}
\newcommand{\Sersic}{S\'{e}rsic}
\newcommand{\degree}{$^{\circ}$}
\renewcommand{\thefootnote}{\fnsymbol{footnote}}


\title{\bf Investigating the Core Morphology--Seyfert Class relationship with Hubble Space Telescope Archival Images of local Seyfert galaxies}
\author{M. J. Rutkowski\altaffilmark{1}, P. R. Hegel\altaffilmark{1}, Hwihyun Kim\altaffilmark{1}, Kazuyuki Tamura\altaffilmark{2}, R. A. Windhorst\altaffilmark{1}}
\altaffiltext{1}{School of Earth and Space Exploration, Arizona State University, Tempe, AZ 85287-1404, USA}
\altaffiltext{2}{Naruto University of Education, Nakashima, Takashima, Naruto-cho, Naruto-shi,772-8502 Japan}

\begin{abstract}

The Unified Model of Active Galactic Nuclei (AGN) has provided a
successful explanation for the observed diversity of AGN in the local
Universe. However, recent analysis of multi\hyphensym wavelength
spectral and image data suggests that the Unified Model is only a
partial theory of AGN, and may need to be augmented to remain
consistent with all observations. Recent studies using high spatial
resolution ground\hyphensym~and space\hyphensym based observations of
local AGN show that Seyfert class and the ``core'' ($r$\,\,\lsim\,\,1
kpc) host\hyphensym galaxy morphology are correlated. Currently, this
relationship has only been established {\it qualitatively}, by visual
inspection of the core morphologies of low redshift ($z<0.035$)
Seyfert host galaxies \citep{M98}. We re\hyphensym establish this
empirical relationship in {\it Hubble Space Telescope} ({\it HST})
optical imaging by {\it visual} inspection of a catalog of 85 local
($D<63$Mpc) Seyfert galaxies. We also attempt to re\hyphensym
establish the core morphology\hyphensym Seyfert class relationship
using an automated, non-parametric technique that combines both
existing classification parameters methods (the adapted CAS,
G\hyphensym M$_{20}$), and a new method which implements the {\tt
  Source~Extractor} software for feature detection in
unsharp\hyphensym mask images. This new method is designed explicitly
to detect dust features in the images. We use our automated approach
to classify the morphology of the AGN cores and determine that Sy2
galaxies visually appear, on average, to have more dust features than
Sy1. With the exception of this ``dustiness'' however, we do not
measure a strong correlation between the dust morphology and the
Seyfert class of the host galaxy using quantitative techniques. We
discuss the implications of these results in the context of the
Unified Model.


\end{abstract}
\section{Introduction}\label{sec:intro}

Active Galactic Nuclei (AGN) are sustained by the accretion of material
from their local environment onto a super\hyphensym massive
(M\,\,\gsim\,\,10$^{6}$-10$^{7}$\Msol) black hole. In the Unified
Model of AGN, the observed diversity in emission-line profiles of AGN
is believed to be an observational bias introduced by the relative
inclinations (with respect to the observer) of the central engine as
it is nested within a toroid of dense molecular material
\citep{B84,A93}. Observations of the ``zoo'' of AGN (e.g., Seyferts,
BL LAC objects, Radio galaxies) from X\hyphensym ray to radio
wavelengths have been remarkably well\hyphensym explained by the
Unified Model \citep[for a review, see e.g.,][]{U95}.

Despite the success of the model, numerous AGN in the local Universe
are not well\hyphensym explained within the paradigm of the Unified
Model. Many tests of the Unified Model have concentrated on the
observed diversity in the properties of Seyfert galaxies, which are
broadly classified by their emission line profiles as$\colon$ a)
Sy1\hyphensym1.9 (Sy1), observed with both broad ({\it
  v}\,\,\gsim\,\,10$^3$km s$^{-1}$) and narrow line emission; and b)
Seyfert 2 (Sy2), observed only with narrow line emission. For example,
\cite{T01,T03} identified Sy2 AGN that lack ``hidden'' Sy1 AGN as
predicted by Unified Model, indicating that Sy2s may not be\longdash
as a class\longdash identical to Sy1 AGN. Furthermore, \cite{PB02}
found that the column density of absorbers in Sy2 AGN implies the
existence of dust absorbers on a larger physical scale
($r$\,\,\gsim\,\,1 kpc) than the molecular toroid. Recently,
\cite{R11} found ``excess'' X\hyphensym ray emission from reflection
in Sy2 AGN, that did not appear to a comparable extent in Sy1 AGN,
indicating an environmental distinction between these two classes of
AGN.

Malkan, Gorjian, \& Tam (1998, hereafter MGT98) tested the Unified
Model via a ``snapshot'' campaign (see \S \ref{sec:thedataset} for
details) conducted with {\it HST} Wide Field Planetary Camera 2
(WFPC2) using the F606W ($\lambda_0$=5907\AA) filter, in which they
observed the morphology of the inner core ($\sim$1 kpc) of 184 local
(z\,\,\lsim\,\,0.035) Seyfert galaxies. The authors visually inspected
these images and determined that Sy1s are preferentially located in
galaxies of ``earlier\hyphensym type'' core morphology, and conversely
that Sy2 AGN are more often hosted by galaxies with ``later\hyphensym
type'' cores. \malkcit~also determined that the distribution of dust
is more irregular and extends closer to the nucleus in Sy2 galaxies
than it does in Sy1 AGN. Hereafter, we refer to these two empirical
relationships as the ``MGT98 relationship.'' These independent studies
suggest that there may be a fundamental physical distinction between
\syall~galaxies, that is not explained by the relative inclination of
the thick, gas-rich toroidal in which the AGN central engine is
embedded. The contemporary debate on the nature of AGN is not framed
exclusively by the Unified Model; other models of the central engine
and the dusty accretion disk do exist \cite[e.g., the ``clumpy torus''
  model of][]{N08}, but we will discuss our analysis in the context of
the Unified Model to provide an easier comparison with published
results in the literature.

In the past two decades, {\it HST} images and spectroscopy has been
used to study hundreds of low redshift ($z<0.1$) AGN, and many of
these observations are now available in the Hubble Legacy Archive
(HLA)\footnote{{\tt http$\colon$//hla.stsci.edu}}, an online
repository maintained as a service for the community by the Space
Telescope Science Institute.

In this study, we test the Unified Model using images downloaded from
the HLA. Specifically, we re\hyphensym examine and extend the analysis
first established in \malkcit~using a catalog of 85 Seyfert galaxies
selected using the criteria outlined in \S \ref{sec:thedataset}. In \S
\ref{sec:visclass}, we present the results of the visual inspection
and classification of our catalog Seyfert galaxies. In \S
\ref{sec:modginim20}, we present, apply and discuss an automated
technique, which we use to quantify the distribution of any dust
features (e.g., dust, stellar clusters, etc.)  present in the cores of
our catalog galaxies. This classification technique quantifies the
distribution of the dust features that were used to qualify the degree
of dust irregularity or morphological class of a galaxy in the
original visual inspection in \S \ref{sec:visclass}. In
\S\ref{sec:sexintro}, we present a new automated technique developed
to detect the dust features, which were identified in
\S\ref{sec:visclass} and used in the visual classification of the
galaxies' cores. We discuss the results, and implications, of our
qualitative visual and quantitative automated analysis in \S
\ref{sec:conclusion}. Throughout, we assume a $\Lambda$CDM cosmology
with $\Omega_m$=0.27, $\Omega_{\Lambda}$=0.73, and $H_0$=70 km$^{-1}$
s$^{-1}$ Mpc$^{-1}$ \citep{K11}.

\section{Data and Image Processing}\label{sec:thedataset}

To test the \malkrel, we require a sufficiently large sample of Sy1
and Sy2 AGN to ensure that any result can be interpreted in a
statistically meaningful way. We therefore use the following selection
criteria to identify this sample of AGN$\colon$

\begin{itemize}

\item {\it Initial Catalog}$\colon$ We develop a large (N$\simeq$240)
  catalog from three large {\it HST} surveys of Seyfert galaxies (Ho et
  al. 1997; MGT98; Ho \& Peng 2001) that were included in the
  NASA/IPAC Extragalactic Database (NED\footnote{available online at
    {\tt http$\colon$//ned.ipac.caltech.edu})}). We refer the reader to
  the respective surveys for specific details associated with the
  sample selection of these AGN. Together, these surveys can be used
  to produce a catalog that is generally representative of the
  morphological diversity of Seyfert galaxies, although none of the
  samples is strictly volume complete.

\item {\it HST WFPC2 F606W HLA Images}$\colon$ At optical wavelengths,
  the resolution of features with a linear spatial extent of
  10\,\,\lsim\,\,{\it r} (pc)\,\,\lsim\,\,100 can only be achieved
  with large\hyphensym aperture space\hyphensym based
  observatories. Thus, we required our galaxies to have {\it HST}
  WFPC2 F606W filter images in the HLA, prepared as mosaics of the
  WF1\hyphensym 3 and PC CCD images multidrizzled\footnote{see
    http$\colon$//stsdas.stsci.edu/multidrizzle/} to a uniform
  0\farcs10 pixel scale. We only used the mosaiced images to ensure
  that the intrinsically different pixel scales of the individual CCDs
  did not bias the identification and classification of sources. The
  HLA contains an image for more than 90\% of the galaxies included in
  our initial catalog with this specific camera and filter
  combination. The fact that these images are available is partly a
  selection bias. Many of the observations we include in our catalog
  were observed by \malkcit~in the snapshot campaign. Note that the
  F606W filter samples longward of the 4000\AA$\,$break at all
  relevant redshifts in our catalog.  This broad filter includes the
  rest-frame H$\alpha$ and [NII] line emission which, in AGN, can be
  prominent. In \S\ref{sec:visclass} we discuss the effect of this
  emission on our qualitative analysis.

\item {\it ``Face\hyphensym On''}$\colon$ We only included
  ``face\hyphensym on'' galaxies to ensure that the dust features
  classified in \S \ref{sec:visclass} are physically confined to a
  region relatively close to the core (1 kpc) of the galaxy. We
  estimated the angle of inclination by eye, and excluded an
  additional 20\% of AGN that appeared at inclinations approximately
  greater than 30\degree. We did {\it not} exclude those galaxies with
  inclination angles that could not be estimated (i.e., irregular
  galaxies), nor do we exclude elliptical galaxies.

\item {\it Distance less than 63 Mpc}$\colon$ We are interested in
  characterizing the structural properties of dust features with a
  linear size scale greater than 100 pc (for more details, see \S
  \ref{subsec:sextractmeth}). We require at least 3.5 WFPC2 pixels
  (0\farcs35 in the HLA mosaic images) to span this physical scale.
  This sets the maximum allowable distance to a catalog galaxy of 63
  Mpc or, equivalently, to a redshift $z$\,\,\lsim\,\,$0.015$.
  Sub\hyphensym kiloparsec scale features (e.g., dust lanes and
  clump\hyphensym cloud formations such as bars, wisps, and tidal
  features like warps and tails) are easily discernible in galaxies
  nearer than this distance observed at the HST spatial resolution. We
  excluded an additional 50\% of galaxies that were at distances
  greater than 63 Mpc. We model and discuss the dependence of the
  morphological classification parameters on spatial resolution
  (Appendix A), and the galaxy distance (Appendix B).

\end{itemize}

\begin{figure}
\centering
\begin{tabular}{c}
  \includegraphics[width=6in,angle=0,scale=1.0]{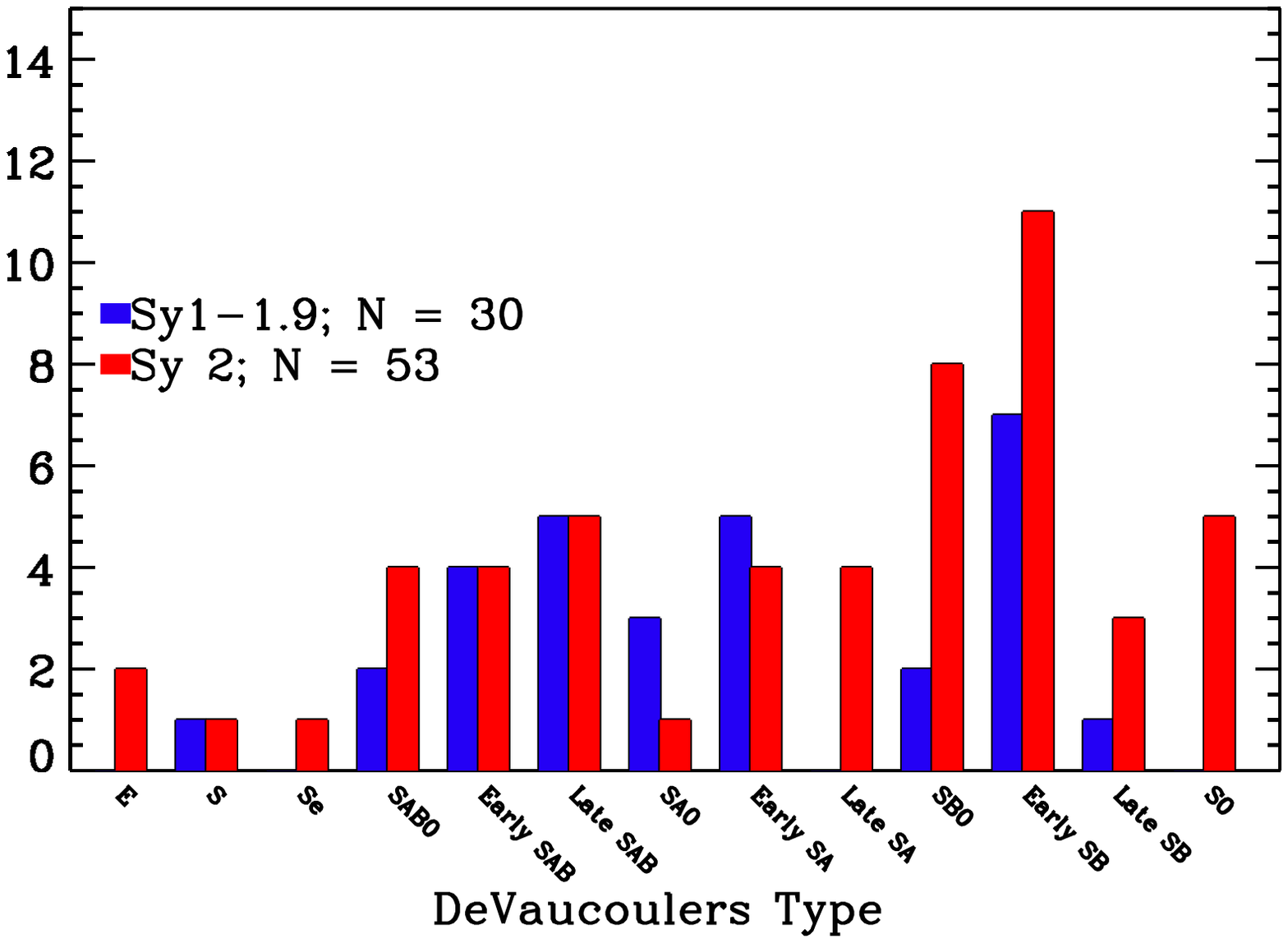} \\
\end{tabular}
\caption{The distribution of galaxy morphologies compiled from NED.
  Two Sy1 AGN were not classified in NED. Though these galaxy
  morphologies are defined for the entire galaxy\longdash {\it not}
  the core region which we are investigating \longdash the similarity
  of these distributions confirms that that any distinction that we
  draw between these classes of AGN is not likely to be attributed to
  the full galaxy morphology.  Furthermore, neither class of AGN is
  biased to a particular class of galaxy, nor are we biased generally
  by our selection criteria towards fundamentally less\hyphensym dusty
  galaxy types (i.e., early\hyphensym type galaxies). }
\label{fig:morphodistro}
\end{figure}

31 Sy1 and 54 Sy2 galaxies from the initial sample met all of these
selection criteria, combined for a total of 85 Seyfert galaxies. This
large sample ensures that the (Poisson) uncertainties from small number
statistics are small.  Our catalog includes significantly fewer Sy1
than Sy2 galaxies, partly due to a bias towards Sy2 AGN in the initial
sample. For example, only 44\% of the galaxies in \malkcit~are
classified as Sy1 AGN. In the Unified Model, this represents a bias in
the opening angle through which the AGN is viewed. Though this bias
may be present, it will not significantly affect this study, because
we are investigating the {\it core} morphological distinctions between
the AGN sub-classes of the host galaxies (i.e., on scales of hundreds
of parsecs, well beyond the $\sim$parsec scale of the thick, dusty
torus).  Where the data are available from NED, we plot the number of
Seyfert galaxies by their deVaucouleurs galaxy type and 60$\mu$m flux
in Figures \ref{fig:morphodistro} \& \ref{fig:oiifir}, respectively.
These figures demonstrate that our catalog is not strongly dominated
by a particular galaxy type or observed AGN luminosity.

We prepared the HLA mosaiced images for analysis by first visually
identifying the (bright) center of each galaxy. We extracted a core
region with physical dimensions of 2$\times$2 kpc centered at this
point. The HLA images that we used have only been processed to the
Level 2 standard, i.e., only images acquired during the same visit are
drizzled and mosaiced in the HLA. Many of the galaxies were originally
imaged as part of {\it HST} snapshot surveys (single exposures with
$t_{exp} \simeq$ 500s). As a result, cosmic rays can be a significant
source of image noise in the mosaics.  We used the routine, $\lacos$
\citep{vdo01} to clean the CCD images of cosmic
rays\footnote{available online at {\tt
    http$\colon$//www.astro.yale.edu/dokkum/lacosmic/}}. Initially, we
implemented $\lacos$~using the author's suggested parameters, but
found by iteration that a lower value for the object\minussign
detection contrast parameter, {\tt sigclip}=2.5, produced cleaner
images without significantly affecting the pixels of apparent
scientific interest. Additional cleaning and preparation of the
imaging was necessary for the following analyses, and we discuss those
task\hyphensym specific steps taken in \S\ref{subsec:methodol}.

\section{Visual Classification of Core Morphology}\label{sec:visclass}

The core morphologies of the AGN\hyphensym host galaxies are diverse
and early\hyphensym~and late\hyphensym type morphologies, with varying
degrees of complexity in dust and gas features, are represented in the
catalog. In Figure \ref{fig:fourpanels}, we provide images of a subset
(12) of our galaxies; each image has been scaled
logarithmically. Images of all (85) galaxies are available in the
electronic supplement to this article.  Here, we use this subset of
galaxies specifically to discuss the various dust features and
structures that we classify by eye.

Galaxies in our catalog display a wide variety of spiral
arms\hyphensym like features. In Figure \ref{fig:fourpanels}, we
provide images of two galaxies (MARK1330, NGC3081) that show distinct
spiral arms. These features are easily distinguished from the ambient
stellar light profile in the core of each image. In some galaxies,
these arms are reminiscent of galactic\hyphensym scale spiral
features, such as barred spirals (NGC3081). Some spiral\hyphensym arm
like features are more unique. For example, MARK1330 has a single arm
that appears to in-spiral and connect with the bright core of the
galaxy. Furthermore, some galaxies appear to be relatively dusty with
numerous features of various size scales, appearing in either
organized or chaotic features (e.g. NGC1068, NGC1386, NGC1672,
NGC3393).

In Figure \ref{fig:fourpanels}, we also provide examples of galaxies
whose cores are relatively sparsely populated with dust features. In
some cases (e.g., NGC3608), these galaxies have few dust features. In
other galaxies (e.g., NGC1058) dust features appear most pronounced in
the core of the galaxy (100-200 pc) and are less significant at large
radii.

We have visually inspected and classified each of the 85 galaxies in
our catalog, first using the following criteria that were defined and
used in \malkcit. We divide these criteria into two general
classes$\colon$

Class 1\longdash Dust Classifiers$\colon$
\begin{itemize}
\item {\bf DI} $\colon$ Irregular dust;
\item {\bf DC} $\colon$ Dust\hyphensym disk/Dust\hyphensym lane passing close or through center (i.e., bi\hyphensym sected nucleus);
\item {\bf D} $\colon$ Direction of dust lanes on one side of major
 axis, where direction is N, S, E, W, NW, NE, SW, or SE;
\item {\bf F/W} $\colon$ Filaments/wisps, and; 
\end{itemize}

Class 2\longdash Ancillary Classifiers$\colon$
\begin{itemize}
\item {\bf R} $\colon$ Ring;
\item {\bf E/S0}$\colon$ Elliptical or Lenticular;
\item {\bf B} $\colon$ Bar;
\item {\bf CL} $\colon$ Cluster, lumpy H {\sc \romannumeral2} region, knots;
\end{itemize}


Four observers (PRH,HK,MJR,KT) inspected the 2$\times$2 kpc postage
stamp images in Figure \ref{fig:fourpanels} and classified each of the
Seyfert cores. We did not use the ``normal'' classifier, because its
definition could not be independently inferred from \malkcit. In
practice, we note that galaxies that showed regular spiral and dust
features in their core morphology were more often classified as
F/W. Conversely those with more irregular spiral and dust features was
classified as DI. These classifications are not mutually exclusive,
i.e., galaxies could be classified as both DI and F/W. In Table
\ref{tab:gallist} all unique visual inspections are provided.



The majority (91\%) of galaxies were identified with dust features.
Irregular dust features (DI) were observed in 42\% (13/31) of Sy1 and
57\% (31/54) of Sy2 AGN. In contrast, 68\% (21/31) of Sy1, and (31/54)
57\% of Sy2 host galaxies, showed regular filaments and wispy features
(F/W).
Thus, by visual inspection, we find that Sy1 host galaxies are
more regular in their dust morphologies than are Sy2 host
galaxies, while Sy2 host galaxies are more chaotic or
irregular in their dust morphologies than are Sy1 host galaxies.

To reduce ambiguity in the classification of regular and irregular
dust features in our galaxies, and to provide a second confirmation of
the \malkrel, we developed an additional system specifically for the
classification of the core dust morphology of Seyfert galaxies. This
classification scheme is defined as follows$\colon$

\begin{itemize}
\item {\bf 1}-``Nuclear spiral''\longdash Distribution of features resembles a flocculent or grand-design spiral;
\item {\bf 2}-``Bar''\longdash A bar\hyphensym like feature in
 emission or absorption extends outward from the center of the
 galaxy;
\item {\bf 3}-``Dust-specific classification''\longdash The previous
  designations considered all structure. The following classifications
  describe only the quality and spatial distribution of what we
  consider to be dust$\colon$
\begin{itemize}
\item {\it Group A}$\colon$
 \begin{itemize}
 \item[--]{\bf s}-``Late-type Spiral''\longdash Dust appears distributed
 in a spiral pattern throughout more than 50\% of the image. The
 ``inner-arm'' regions appear to be clear of any dust;
 \item[--]{\bf i}-``Irregular''\longdash No visually distinguishable
   pattern can be identified in the spatial distribution of dust,
   i.e., the dust is patchy and irregular in form;
 \end{itemize}
\item {\it Group B}$\colon$
 \begin{itemize}
 \item[--]{\bf m}-``High Extinction''\longdash Dust
 features appear to be of high column density. The galaxy appears
 highly extincted. Dust lanes appear to ``cut'' through the ambient
 stellar light of the galaxy;
 \item[--]{\bf l}-``Low Extinction''\longdash Low contrast dust is
 present, but is barely discernible from the ambient stellar light.
 \end{itemize}
\end{itemize}

\end{itemize}

In Table \ref{tab:fitpars}, we provide the classification using this
scheme. If possible, galaxies were classified using Class 1 and 2, but
all galaxies were classified according to their dust structure (Class
3). The sub-groups of Class 3 (A\&B) were mutually exclusive; e.g., no
galaxy could be classified as `3is'.  Galaxies could be classified by
a single Group A and one Group B classification simultaneously (e.g.,
`3mi'). If there was a conflict in classifying dust structure amongst
the four co-authors, the {\it majority} classification is listed in
Table \ref{tab:fitpars}.  If no majority was reached after first
classification, the corresponding author made the final classification
{\it without knowledge of the Seyfert class} in order to prevent any
unintentional bias in the measurement of the Malkan relationship.

The WFPC2 F606W filter we used in this image classification is broad
($\lambda\lambda\simeq$4800-7200\AA) and includes the H$\alpha+$[NII]
line complex.  In principle, this line emission could affect our
visual classification. In practice, the contribution of {\it line
  flux} to the continuum is relatively minor --- the contribution of
the [NII] doublet to the total flux in this bandpass using the SDSS
QSO composite spectrum \citep{vdb01} is $\ll$1\%, and we estimate the
ratio of the equivalent widths, EW$_{\mbox{H}\alpha}$/EW$_{[NII]}$, of
these lines to be $\simeq\!3\colon\!2$.  Despite the minor
contribution to the total observed flux in line emission, the
photo\hyphensym ionization of the gas-rich local medium by the central
engine can produce significant ``hotspots'' at the wavelengths of
these atomic lines, which appear as structure in the
image. \cite{Co00} has studied an example of this photo\hyphensym
ionization structure, the spiral-like ``S'' structure in one of our
sample Seyferts hosts (NGC3393; see Figure
\ref{fig:fourpanels}). Though this emission contributes very little to
the total {\it flux} in the core, the high contrast between these
bright emitting sources and the local area could lead to ``false
positive'' classifications of dust features. Fortunately, few AGN
($\sim$7-8 galaxies, see e.g., MARK3, MARK1066, NGC1068, NGC3393,
NGC4939, \& NGC7682 in Figure \ref{fig:fourpanels} of the online
supplement) show evidence of these emitting structures and these
highly localized structures are easy to distinguish, in practice, from
the stellar and dust continuum.

In  conclusion, we confirm  that Sy2  host galaxies  are significantly
{\it more} likely to  have irregular core morphologies$\colon$ 58\% of
Sy2 host galaxies  were classified as `3i'. In  contrast, only 40\% of
Sy1 host galaxies were classified  as `3i'.  Furthermore, 39\% Sy2 AGN
were classified as `3s' in contrast  to 53\% of Sy1 host galaxies. The
results of  our visual classification agrees with  the observations in
\malkcit.

Although visual inspection can be an effective means of classifying
the morphology of spatially resolved sub\hyphensym structure in
galaxies, it has its disadvantages as well.  Visual inspection is
time\hyphensym consuming, it does not provide a quantifiable and
independently reproducible measure of the irregularity of structures
that can be directly compared with the results of similar studies, and
it can be highly subjective.  Though guidance was provided to the
co-authors on how to classify varying degrees of dust structure using
the Class 3, such classifications are highly subjective and conflicts
in classification could arise between co-authors.  For example,
approximately 55\% of the visual classifications of dust structure
(Table \ref{tab:fitpars}) were {\it not} unanimous.  This discrepancy
can be largely attributed to the subjective definition of the Class 3
sub-classifications.  In each galaxy, the co-authors implicitly
emphasized certain dust features when making their classification.  In
many galaxies, whether the authors chose to weight the significance of
physically small or large-scale dust structure could change the
structural classification significantly.  Consider the case of
NGC1365$\colon$ this galaxy was classified with an irregular dust
morphology due to the small-scale dust features that appear to
dominate the visible sub-structure in the core.  But, authors who
(subconsciously or otherwise) emphasized the broad dust ``lanes'' in
the north and (to a lesser extent) south may classify the core as
having a ``spiral'' dust morphology. Neither classification is
necessarily incorrect --- the broad dust lanes are clearly associated
with the prominent spiral arms in this galaxy when viewed in full
scale.  These complicating factors can weaken any conclusion drawn
from the visual classification of galaxies.

In recent decades, as image analysis software and parametric
classification techniques have become prevalent, the astrophysical
community has implemented automated methods for galaxy classification
\citep[e.g.,][]{O96,C03,L04}. By relegating the task of object
classification to automated software and algorithmic batch processing,
these methods have gained popularity, because they can significantly
reduce the time observers must spend inspecting each galaxy, and can
provide a reproducible classification for each galaxy
\citep[cf.][]{Li08}.

Therefore, we extend our original test of the Unified Model to include
a quantitative assessment of the morphological differences between
Seyfert galaxies that we identified by visual inspection. We present
these techniques in \S\ref{sec:modginim20} and
\S\ref{sec:sexintro}. Our use of quantitative parameters to measure
morphological distinctions between Seyfert core morphologies can
provide a more robust test of the \malkrel$\,$ by reducing some of the
biases implicit in visual inspection.

\section{Traditional Quantitative Morphological Parameters}\label{sec:modginim20}
A variety of parameters have been defined in the literature to
quantify galaxy morphology. These parameters are distinguished by
their use of a pre-defined functional form\longdash i.e., parametric
or non-parametric\longdash to express galaxy morphology. Some popular
non-parametric morphological parameters are ``CAS''
\cite[][``Concentration'', ``Asymmetry'', and ``clumpinesS'']{C03} and
``Gini\hyphensym M$_{20}$'' \citep[][``Gini Coefficient'' and
  M$_{20}$, the second\hyphensym order moment of brightest 20\% of the
  galaxy pixels]{A03,L04}. These methods are not without limitations
\cite[cf.][]{Li08}, but they each can be useful for assessing galaxy
morphology in a user\hyphensym independent and quantitative way.  We
chose to use these parameters in this research project, because the
distribution of dust features in the cores of Seyfert galaxies is
unlikely to be well-described by a single functional form, e.g., the
S\'{e}rsic function that broadly distinguishes between elliptical and
spiral galaxy light profiles.

\cite{C00} provide the following functional definitions of the CAS
parameters.

The concentration index, C, is defined as$\colon$
\begin{equation}\label{eqn:CO}
 {\mbox C}=5~{\mbox ln}~\left(\frac{r_{80}}{r_{20}}\right),
\end{equation}
where {\it r}$_{80}$ and {\it r}$_{20}$ are the values of the circular
radii enclosing 80\% and 20\% of the total flux. The typical range in
concentration index values measured for galaxies on the Hubble
sequence is $1$\,\,\lsim\,\,C\,\,\lsim\,\,$5$ \citep{C04,H08}. Larger
values of the concentration parameter are measured for galaxies that
are more centrally peaked in their light profiles.

 The asymmetry, A, is defined as$\colon$
\begin{equation}\label{eqn:AS}
{\mbox A}=\frac{\displaystyle\sum\limits_{a,b=0}^{x,y}|I_o(a,b)-I_\Phi(a,b)|}{2\sum\limits_{a,b=0}^{x,y}|I_o(a,b)|},
\end{equation}
where {\it x} and {\it y} correspond to the length (in pixels) of the
image axes, $I_o$ is the original image intensity, and $I_{\Phi}$ is
intensity of pixels in an image that is the original image rotated
through an angle of $\Phi$ (we set $\Phi=180$\degree). Typically, A
ranges from 0 (radially symmetric) to 1 (asymmetric), see e.g.,
\cite{C03}.

 Clumpiness, S, is defined as$\colon$
\begin{equation}\label{eqn:CL}
{\mbox S}=10 \times\sum\limits^{x,y}_{a,b=0} \frac{(I_o(a,b)-I^{\sigma}(a,b))-B(a,b)}{I_o(a,b)},
\end{equation}
where $I_o(a,b)$ is the image intensity in pixel (a,b),
$I^{\sigma}(a,b)$ is the pixel intensity in the image convolved
with a filter of Gaussian width $\sigma$, and B(a,b) is the estimated
sky\hyphensym background for a given pixel. Typically,
0\,\,\lsim\,\,S\,\,\lsim\,\,1 \citep[see e.g.,][]{C03}, and galaxies
that appear to be visually ``clumpier'' have higher values of S.

\cite{A03} and \cite{L04} provide the following functional definitions
of the Gini\hyphensym M$_{20}$ parameters. The Gini parameter is
defined as $\colon$
\begin{equation}\label{eqn:GIN}
{\mbox G}=\frac{1}{\bar{f}n(n-1)}\sum_{\it j}^{\it n}(2{\it j}-n-1)f_{\it j},
\end{equation}
where {\it \={f}} is the mean over all pixel flux values {\it f$_{\it
 j}$}, and {\it n} is the number of pixels. This parameter measures
inequality in a population using the ratio of the area {\it between} the
Lorentz curve, defined as$\colon$
\begin{equation}
L(p)=\frac{1}{\bar{f}}\int_{0}^{p}F^{-1}(u)du,
\end{equation} 
and the area under the curve of uniform equality ($=\frac{1}{2}$ of
the total area). Although this parameter was originally developed by
economists to study wealth distribution, this parameter can be applied
to understand the distribution of light in galaxies. If the
distribution of light in galaxies is sequestered in relatively few
bright pixels, the Gini coefficient approximately equals unity. The
Gini coefficient is approximately equal to zero in galaxies in which
the flux associated with each pixel is nearly equal amongst all
pixels. In other words, the Gini coefficient quantifies how sharply
peaked, or ``delta$-$function''\minussign like the flux in galaxies
is.  Note that this parameter can be affected by the ``sky'' surface
brightness estimate assumed by the user, which we discuss in Appendix
C.

The M$_{\mbox{20}}$ parameter is calculated with respect to the total
second\hyphensym order moment, M$_{\mbox{tot}}$, flux per pixel, $f_{\it j}$,
which is defined as$\colon$
\begin{equation}\label{eqn:M20}
{\mbox M_{\mbox{tot}}}=\sum_{\it j}^{\it n}M_{\it j}=\sum_{\it j}^{\it n}f_{\it j}[(x_{\it j}-x_c)^2+(y_{\it j}-y_c)^2],
\end{equation}
such that$\colon$
\begin{equation}
{\mbox M_{20}}=log \left( \frac{\sum_{\it j}^{\it n}M_{\it j}}{M_{tot}} \right), \mbox{while} \sum_{\it j}^{\it n} f_{\it j}<0.2f_{tot}, 
\end{equation}
where M$_{\it j}$ is the second\hyphensym order moment at a pixel
${\it j}$, and ($x_c$,$y_c$) are the coordinates of the central
pixel. In general, M$_{20}$ typically ranges between
$-3$\,\,\lsim\,\,M$_{20}$\,\,\lsim\,\,0 \citep{L04,Lo08,H11b}. If
considered jointly with the Gini coefficient, \cite{L04} determined
that larger values of M$_{20}$ (with correspondingly smaller values of
G) are associated with ``multiple ULIRG'' galaxies, and that M$_{20}$
is a better discriminant of merger signatures in galaxies.

We measure these five parameters\longdash CAS and Gini\hyphensym
M$_{20}$\longdash to quantify distinctions between the distribution of
light, which underpins the classifications we first made in
(\S\ref{sec:visclass}).

\subsection{Case\hyphensym specific Implementation of Traditional Morphological Parameters}\label{subsec:methodol}

The authors of CAS and G\hyphensym M$_{20}$
\citep[][respectively]{C03,A03,L04} each defined a method to prepare
images for analysis that accounts for systematic issues (e.g.,
compensating for bright or saturated cores of the galaxies). This
method of image preparation and analysis also ensures that the
parameters are measured for the galaxy itself, and that the
contributions from background emission are minimized. In our analysis,
we calculate all morphological parameters applying a functional form
that is consistent with\longdash or identical to\longdash the form
presented in the literature. However, we caution that our images and
specific science goals require us to use an algorithm for image
preparation and parameter measurement that differs slightly from the
published methods. In this section, we outline key differences between
our data and methods we used and those presented in the literature.

First, we measured these traditional parameters in images of galaxies
observed at fundamentally different spatial resolutions (see
\S\ref{subsec:origanaly}). All galaxies in our catalog have been
observed with {\it HST} WFPC2 at a pixel scale of 0$\farcs$10
pix$^{-1}$. In contrast, CAS and \ginm~are often measured from images
obtained with ground\hyphensym based telescopes that have relatively
low spatial resolution in comparison with {\it HST}. For example,
\cite{F96} present images obtained with the Lowell 1.1 and Palomar 1.5
meter telescopes at $\sim$ 2\farcs0 resolution at full\hyphensym width
half maximum (FWHM). This data set has been used extensively to test
the CAS and \ginm~parameters' ability to discriminate between the
morphological classes and star\hyphensym formation histories of nearby
galaxies \citep[e.g.,][]{C03,L04,H06,H08}. The vastly different
spatial resolutions between images obtained at ground\hyphensym based
observatories and the {\it HST} images implies that the parameters
that we measure are sensitive to features of fundamentally different
size\minussign scales. In fact, in ground-based images most of the
small\hyphensym scale structure that we identified and used to
classify the galaxies (\S\ref{sec:visclass}) is undetected.  Thus,
parameters that are dependent on the pixel\hyphensym specific flux
values (e.g., M$_{20}$), rather than on the average light distribution
(e.g., concentration index), may be more sensitive to these
spatial\minussign resolution differences because at lower resolution
fine\hyphensym scale structure are effectively smoothed out.  In
Appendix A, we quantify the effect of spatial resolution on the
specific criteria that we use to characterize the structure of dust
features in our Seyfert galaxies.


 In \cite{C03} and \cite{L04}, the CAS and \ginm~parameters,
 respectively, are all measured in an image that is truncated at the
 Petrosian radius of the galaxy. The Petrosian radius is defined as
 the radius ($r_{p}$) at which the ratio of the surface brightness
 {\it at} $r_p$ to the mean surface brightness of the galaxy {\it
   interior} to $r_p$ equals to a fixed value, $\eta$, typically equal
 to 0.2. A Petrosian radius or similar physical constraint is applied
 to differentiate between galaxy and sky pixels, and ensures that the
 influence of background emission is minimized in the calculation of
 CAS and G-M$_{20}$. The mean Petrosian radius measured in the
 r$^{\prime}$\,\hyphensym filter ($\lambda_0$=6166\AA) of Sloan
 Digital Sky Survey (SDSS) Data Release 7 is $\sim$4.3
 kpc\footnote{Only 28 galaxies in our catalog were observed in SDSS
   DR7, available online at {\tt http://www.sdss.org/dr7}, but those
   galaxies common to our survey and SDSS span a range of morphologies
   and distances, hence we consider the measured mean Petrosian radius
   to be representative for our catalog.}. Since we are not interested
 in the contribution of dust features located at radii greater than 1
 kpc, we do not use images truncated at r$_p$. Furthermore, at the
 mean redshift of our sample, the WFPC2 PC chip field of view is
 $\lesssim$ 2.8 kpc.

Unlike observations of the entire galaxy, in our core images we can
make the reasonable assumption that {\it most} of the flux that we
observe in the core arises from sources or features physically
associated with the galaxy. Not all pixels are sensitive to the flux
arising from the galaxy, though, and we use the following method to
differentiate between the light arising from galaxy sources and other
extraneous objects or noise.

\begin{itemize}
\item We set all pixels that occur at the edges and the chip gaps
  between the WFC and PC CCDs in the mosaiced images equal to
  zero. Furthermore, the center of the galaxy is often much
  (10\hyphensym 100$\times$) brighter than the rest of the galaxy,
  likely due to the AGN emission. To avoid such extremely bright
  pixels from biasing the measurement of any of the automated
  classification parameters, we set a high threshold defined as the
  average of the inner\hyphensym most 5$\times$5 pixels for each
  galaxy. We set the pixel values above this threshold equal to zero
  in the CAS \& G\hyphensym M$_{20}$ computations.

\item If the functional form of a parameter explicitly required a
  background term, we set this term equal to zero. Our analysis is
  focused on the cores of each galaxy ($\sim$1 kpc; or less than
  0.5$\times r_p$), which are significantly brighter, and have high
  enough surface brightness, that the contribution of background
  objects can be considered to be minimal. We assume that our images
  include only light from the galaxy itself and background emission
  from the zodiacal (foreground) light, which arises from sunlight
  scattered off of $\sim$100$\mu$m dust grains. Fortunately, from the
  generally dark {\it HST} on-orbit sky, the actual zodiacal sky
  surface brightness is a simple well\hyphensym known function of
  ecliptic latitude and longitude ($\ell^{Eq.},b^{Eq.}$). We use
  measurements of the zodiacal background from WFPC2 archival images
  presented by Windhorst et al. (in prep.), to estimate the emission
  from this dust in the F606W band.  The average on\hyphensym board
  {\it HST} F606W\hyphensym band zodiacal sky brightness can be found
  in Table 6.3 of the WFCP2 Handbook \cite{M08}, but the values in
  Figure \ref{fig:zodiplot} give a more accurate mapping as a function
  of $\ell^{Eq.}$ \& $b^{Eq.}$. The zodiacal background could not be
  directly calculated from the images, because the galaxy core
  typically over\hyphensym filled the CCD. We correct for the zodiacal
  foreground emission prior to image analysis in
  \S\ref{subsec:origanaly} and \S\ref{subsec:sextractmeth}. For more
  details, see Appendix C.

\item To measure clumpiness, we included an additional processing step
  motivated by the algorithm defined in \cite{H11}. Prior to
  calculating the clumpiness parameter as defined in \cite{C03}, we
  first applied a 5$\times$5 pixel boxcar smoothing to the input
  image. We produced the residual map by subtracting the smoothed
  galaxy image from the original input image. The additional smoothed
  image was generated by applying a boxcar smoothing kernel with the
  one\hyphensym dimensional size of kernel defined as$\colon$
  $2.0\times\frac{1}{6}\times\ell$, where $\ell$ is the dimension of
  the galaxy image in pixels. By design (see \S\ref{sec:thedataset}),
  the linear size of the smoothing kernel is equivalent to
  $\frac{4}{6}$ or $\sim$0.67 kpc. If we assume that 4 kpc is
  approximately equal to the Petrosian radius for each galaxy in the
  sample, then this dimension is comparable to the smoothing kernel
  size applied in \cite{C03} and \cite{H11}. We tested this assumption
  of an average Petrosian radius, and found that using a larger or
  smaller value ($\Delta\!=\!\pm2$kpc) for the linear dimension of the
  kernel has less than $\sim$1\% effect on the measurement of
  clumpiness.  In this analysis, we also set all pixels within
  1\farcs0 of the galaxy center equal to zero.

\end{itemize}

In the subsequent analysis, we removed all zero\hyphensym valued
pixels to prevent those pixels from affecting the calculation of any
of the parameters.

Though we use identical\longdash or nearly identical\longdash
functional definitions of each morphological parameter used in the
literature, we are analyzing regions of our galaxies at size\hyphensym
scales that are significantly different than have been used in
previous research. As a result, we cannot assume that our parameter
measurements are directly comparable to the CAS and G$-$M$_{20}$
traditional measurements in the literature
\citep[e.g.,][]{C03,L04}. We therefore refer to our parameters that we
derived using the above criteria hereafter as $C^*,A^*, S^*$ and
$G^*$\hyphensym$M_{20}^*$, in order to distinguish our measurements
from the traditional parameters presented in the literature.

\subsection{Analytical Results and Discussion}\label{subsec:origanaly}

Figure \ref{fig:CGM20plot} provides three permutations of the measured
$G^*$\hyphensym $M^*_{20}$\hyphensym $C^*$ parameters. In this figure,
Sy1 and Sy2 host galaxies are represented in blue and red,
respectively. We use this color scheme for all figures provided in the
online version of this manuscript to distinguish our measurements for
the two classes of Seyfert galaxies.  It is noteworthy that the
distribution of each of these parameters spans a range that is
comparable to the range of the G, M$_{20}$, and C measured from
ground-based images at the lower spatial resolution
(0.7\,\,\lsim\,\,$G^*$\,\,\lsim\,\,0.1,
$-$2.5\,\,\lsim\,\,$M^*_{20}$\,\,\lsim\,\,$-0.5$,
2.5\,\,\lsim\,\,$C^*$\,\,\lsim\,\,5.5).

In Figure \ref{fig:CGM20plot}(a) we overplot a dashed line that
differentiates ``normal'' galaxies (which reside below this line) from
starburst galaxies or ULIRGs (i.e., Ultra Luminous Infrared Galaxies)
as defined in \cite{L04}. Four of our Seyfert galaxies are measured to
be on or above this line$\colon$ NGC1672, NGC4303, NGC4395,
NGC7469. The fact that these galaxies reside in this parameter space
is appropriate, since these four galaxies are considered to be
starburst or circum-nuclear starburst galaxies in the
literature. However, approximately 32\% of the Seyfert galaxies were
identified in the literature as starburst or circum\hyphensym nuclear
starburst galaxies. Hence, we conclude that $G^*$ and $M^*_{20}$ do
not effectively discriminate between ``normal'' and starburst
galaxies, as these parameters are demonstrated to do in the
literature. We note that $G$-$M_{20}$ are used to distinguish
starburst and ``normal'' galaxies when the complete galaxy morphology
is considered.  The morphology of the galaxy on this scale need not
necessarily match with the core morphology of the galaxies.

We can consider the {\it relative} distribution of the
$G^*$\hyphensym$M^*_{20}$ values measured for our AGN. In Figure
\ref{fig:CGM20plot}(a), we fit a Gaussian function to the $G^*$ and
$M^*_{20}$ distribution and measure the shape, centroid, and peak of
this function for both Sy1 and Sy2 AGN to be comparable. The
parameters of the fitted Gaussian function are provided in Table
\ref{tab:spartab1}.

We can draw similar conclusions as above from the distribution of
$M^*_{20}-C^*$ and $C^*-G^*$ presented in Figure
\ref{fig:CGM20plot}(b) and (c), respectively. First, it is noteworthy
that $C^*$ is well\hyphensym distributed in the same parameter space
spanned by the conventional concentration index, when it was
calculated for the entire galaxy at lower spatial resolution. We fit a
Gaussian to the $C^*$ distribution measured for Sy1 and Sy2 AGN, and
measured comparable values for the centroid and FWHM of each
distribution (see Table \ref{tab:spartab1}).

We perform a two\hyphensym sample Kolmogorov\hyphensym Smirnov
(K\hyphensym S) for the Sy1 and Sy2 distributions to test whether
these distributions are self\hyphensym similar. The two\hyphensym
sample K\hyphensym S test can be used to measure the likelihood that
two empirical distributions were drawn as independent samples from the
same parent distribution. We use the K\hyphensym S test here for two
reasons, in contrast to more commonly measured statistical parameters
(e.g., the $\chi^2$ statistic)$\colon$ 1) the sample size for each
distribution is small, which can lead to an incomplete distribution;
and 2) {\it a priori} we do not know the parent distributions from
which the empirical distributions were drawn. We use the IDL routine
{\tt kstwo} to measure the K\hyphensym S statistic, {\it d}, which
equals to the supremum distance between the cumulative distribution
functions (CDF) of the input distributions. {\tt kstwo} also reports
the probability statistic, {\it p}, which is the likelihood of
measuring the same supremum in a random re-sampling of the parent
distributions expressed by the empirical distributions. The
K\hyphensym S test cannot provide any insight into the parent
distribution(s) from which the empirical distributions are drawn, but
it can be used to test the null hypothesis that the empirical
distributions were drawn from the same parent distribution. When the
K\hyphensym S statistic is small or the probability is large
(p$>$0.05), the null hypothesis cannot be rejected with confidence.
 
The results of the K\hyphensym S test for the $M^*_{20}$ and $G^*$
parameter distributions are provided in Table
\ref{tab:spartab1}. These distributions are indistinguishable for both
Seyfert classes.  However the K\hyphensym S test measures a slightly
larger values of {\it d=}0.38 for the distribution of C$^*$,
indicating that the CDFs are distinct. The associated probability
statistic for $C^*$ is small ({\it p=}0.01). We conclude that the
$C^*$ distributions measured for \syall~are significantly different,
and thus are likely to be drawn from unique independent parent
distributions. This could support the morphological distinction
between the cores of Sy1 and Sy2 galaxies that was identified by
visual inspection in \S\ref{sec:visclass}. In contrast, if
$G^*$\hyphensym$M^*_{20}$ are indeed sufficiently robust metrics for
distinguishing the distribution of light in the cores of these Seyfert
galaxies, then the results of the K\hyphensym S test suggest that
these parameters do not quantitatively distinguish the galaxy
morphologies of Sy1 and Sy2 AGN.

We consider the $A^*$ (asymmetry) and $S^*$ (clumpiness) parameters
independently from the $G^*$\hyphensym$M^*_{20}$ parameters, because
we believe that these parameters are best able to identify\longdash by
design\longdash the dust features that we found by visual
inspection. In Figure \ref{fig:asymdist}(a), we present the
distribution of $A^*$ measured for our AGN. We did not calculate
asymmetry for NGC1058, NGC1386, NGC1672, NGC3486, NGC4051, NGC4303,
NGC4395, and NGC4698, because the WFPC2 images of these galaxies
included off\hyphensym chip regions that were set to zero (see \S
\ref{subsec:methodol}). These regions can seriously affect our
measurements because asymmetry is calculated by differencing a rotated
image with the original. In Figure \ref{fig:asymdist}(b), we present
the distribution of $A^*$\&$S^*$.  We fit Gaussian functions to the
distributions of $A^*$ and $S^*$ measured for \syall,
respectively. The best\hyphensym fit Gaussian function to each
distribution are provided in Table \ref{tab:spartab1}.  The Gaussians'
parameters measured for \syall~galaxies appear to be
indistinguishable. We confirm this via a two-sample K\hyphensym S test
for the $A^*$ and $S^*$ distributions.  The results of this test are
presented in Table \ref{tab:spartab1}. We conclude from this test that
both parameters distributions are likely drawn from the same parent
distribution.

The uniformity in the $C^*, A^*, S^*,$ \& G$^*$-M$^*_{20}$ distributions
also suggests that the H$\alpha+$[NII] emission arising from the
photo\hyphensym ionization of gas (see \S\ref{sec:visclass}) does not
strongly affect the measurement of these parameters.  Furthermore, if
the $A^*$ and $S^*$ parameters are suitable metrics for quantifying
the morphology of galaxies, then the results of this quantitative
analysis do not support the correlation between core dust morphology
and Seyfert class established by \malkcit~and confirmed by our visual
inspection in \S \ref{sec:visclass}.

In conclusion, four of the five quantitative parameters ($A^*,S^*,
G^*,$ and $M^*_{20}$) measured for the galaxies do not support the
qualitative conclusions developed from visual inspection. The
distribution of $C^*$ may be specific to the class of AGN, which could
support the \malkrel, but this parameter is the least\hyphensym
suited, by design, to quantify the morphological distinctions that
supported the morphology\hyphensym AGN class correlation.  We note
that this does imply that the \malkrel~ is invalid, but that the use
of the $C^*,A^*,S^*,$ \& $G^*$\hyphensym$M^*_{20}$ parameters to
confirm this distinction may not be optimal.

\section{Quantitative Morphology with Source Extractor}\label{sec:sexintro}

The results of the previous analysis may imply that $C^*,A^*,S^*,$ \&
$G^*$\,\hyphensym$M^*_{20}$ parameters are insufficient as tools to
distinguish the sub\hyphensym kiloparsec scale features in AGN.  In
this section, we therefore develop an additional non\hyphensym
parametric technique that uses {\tt Source~Extractor}
\cite[][hereafter, \sex]{B96} to measure the distribution of dust
features in the cores of AGN host galaxies.

\sex~is an automated object detection software package that generates
photometric object catalogs.  This software is widely used for
photometry and star/galaxy separation in UV\hyphensym
optical\hyphensym IR images partly due to the software's speed when
applied to large image mosaics. A review of the literature returns
more than 3000 citations to \cite{B96}, with applications extending
even beyond astrophysics \citep[e.g., medical imaging of tissue
  cultures by][]{T10}. The versatility of \sex~to detect and measure
the properties (e.g., aperture photometry) of galaxies motivated us to
adapt \sex~for our purposes. In our study, we use \sex~only for object
detection, because the algorithm we outline
(\S\ref{subsec:sextractmeth}) and apply (\S\ref{subsec:sexresults})
may prevent accurate photometry.

\sex~has often been used in the study of nearby, dusty galaxies
\citep[see recent work by][for example]{K12,H12}.  This research does
not employ \sex~to directly detect and measure the properties of the
absorbers.  Rather, \sex~is used to derive the photometric properties
of galaxies, and these data are coupled with the dust properties of
the galaxy (e.g., covering fraction). In \S \ref{subsec:sextractmeth},
we adapt \sex~to directly detect dust features that are visible to the
eye. Thus, our use of \sex~to outline the characteristics of dust
features that are fundamentally seen {\it in absorption} is a unique
application of this software.

\subsection{Technical Implementation to Identify Dust Features}\label{subsec:sextractmeth}

In this Section, we outline the method we used to train \sex~to
identify ``objects'' that we visually identified as dust features,
effectively using this software to mimic the human eye.

To detect these objects, \sex~first calculates a local background, and
determines whether the flux in each pixel is above a user\hyphensym
defined threshold {\tt detect\_thresh}. All pixels exceeding this
threshold are grouped with contiguous pixels that also exceed this
threshold. When a sufficient number (defined by the {\tt
  detect\_minarea} parameter) of contiguous pixels are found to meet
the signal threshhold, the pixel group is recorded as an object in the
object catalog. Finally, \sex~measures a variety of parameters (e.g.,
object center, total flux, size, orientation), and constructs a
segmentation map of detected objects.

To detect objects corresponding to the visually detected dust features
in the cores of our galaxies, it was necessary to first train
\sex~using the WFPC2 images of the Seyfert host galaxies. Initially,
we used the HLA image of each galaxy\longdash appropriately cleaned of
defects as detailed in \S\ref{sec:thedataset}\longdash for object
detection. After extensive testing, we could not determine a suitable
combination of the parameters {\tt detect\_minarea} and {\tt
  detect\_thresh} that would force \sex~to identify a set of
comparable objects to the set of dust features that we visually
identified in \S\ref{sec:visclass}. By setting {\tt detect\_thresh}
low enough that nearly all visually identified dust features are
recovered, too many of these features were broken into multiple unique
objects. To alleviate the over\hyphensym segmentation, we increased
the {\tt detect\_minarea} parameter. In order to recover the majority
of the visually identified dust features though, this parameter must
be set unfavorably high; dust features were only detected when they
were included as a component of a much larger, brighter object.

Direct detection of dust features with \sex~is difficult. This can be
directly attributed to the manner in which \sex~detects
objects. \sex~is designed to detect peaks {\it above} the local
background. In the images, the local background is bright, and not
likely to be smooth because it arises from the ambient stellar
background and not the astronomical/zodiacal sky.  Furthermore,
\sex~can not detect many of the dust features as they are observed
    {\it in absorption} with respect to the background. These
    absorption features may be brighter than the true astrophysical
    background, but they are still fainter than the local
    background. Thus, no optimal \sex~parameters can be defined to
    exclusively select the visually identified dust features.

We therefore trained \sex~to identify objects that more closely
matched with dust features identified (\S\ref{sec:visclass}) by
coupling object detection using \sex~with the ``unsharp\hyphensym
mask'' technique. The unsharp\hyphensym mask is a common tool for
image analysis, because it enhances features of specific spatial
scales. In astronomical images, these features correspond to physical
objects, such as stars, star clusters, and/or dust clouds. To apply
this procedure, we first convolved the WFPC2 images with a Gaussian
convolution kernel to create a smoothed image. Next, we divided the
convolved image by the original image to produce the {\it inverse}
unsharp\hyphensym mask image (hereafter, IUM)\footnote{The
  unsharp\hyphensym mask image is typically produced by either
  differencing or dividing the original image by the convolved
  image. When the contrast between the original and the convolved
  image is small, as it is in our WFPC2 images, these two different
  calculations yield similar results.}. In principle, if we
appropriately define our convolution kernel such that it enhances
these structures of specific size\hyphensym scales corresponding to
dust features and apply the IUM, those features should now be detected
as a positive signal above the local background using \sex~with the
appropriate detection parameters. In Figure 6, we provide an
illustration of this technique. In Figure 6a\&6b) we show the core
image of NGC3081 and a surface map of an inter\hyphensym arm
region. We convolved the image with a kernel (Figure 6c), and apply
the IUM technique to produce Figure 6d. In this figure, it is apparent
that the dust features in the region of interest have been enhanced by
the IUM technique.

To produce the IUM image of each galaxy, we first assumed that giant
molecular clouds (GMCs) are physically associated with dust features.
We generated a Gaussian convolution kernel that is specific to each
galaxy, the properties of which were motivated by observations of
GMCs. To produce the appropriate convolution kernel, we first used the
galaxy's redshift from NED to define a physical pixel scale ($s_{p}$;
pixel kpc$^{-1}$) of the kernel. The linear size scale of GMCs is
typically less than 100 pc \citep[see][]{C84,F10}, so we defined the
FWHM of the kernel equal to $\ell$/$s_p$, initially with $\ell$=100
pc. We tested a range of size scales, and determined that $\ell$=80 pc
optimally enhanced the sub\hyphensym kiloparsec scale dust features
that we visually identified in \S\ref{sec:visclass}. We also
determined the appropriate linear size of the kernel to be equal to
{\it x}/10, where {\it x} is the length on each axis of the image in
pixels.

We determined optimal \sex~parameters by an iterative process to find
the segmentation map that most faithfully reproduced the dust features
classified in \S\ref{sec:visclass}. In this process, we fixed the
\sex~parameter {\tt detect\_minarea} equal to 90.0/$s_{p}$ for all
objects. We required {\tt detect\_thresh} for each object pixel to be
at least 1.5$\sigma$ above the local sky\hyphensym background in the
IUM image. 
 Additionally, we determined that the default values for the
 \sex~parameters {\tt deblend\_nthresh \mbox{and} deblend\_mincont}
 equal to 32, and 0.03, respectively, were sufficient for dust feature
 detection in the IUM image.

We discuss the results of implementing our method using the optimized
parameters in \S\ref{subsec:sexresults}. The algorithm we have
outlined above for the detection of dust features in absorption in
images is generic. It is not applicable exclusively to our specific
scientific interests. Thus, we prepared all IDL procedures that we
developed to implement this technique for the public. Readers who wish
to apply this method to other science topics are encouraged to email
the corresponding author.

\subsection{Results and Discussion}\label{subsec:sexresults}

In Figure \ref{fig:fourpanels}, we presented a four\hyphensym panel
mosaic of 12 galaxies including the WFPC2 galaxy core image and its
corresponding \sex~segmentation map. The first of these images was
discussed in \S\ref{sec:visclass}.  
 To produce the second image, we reproduced the segmentation
images in {\tt DS9} using the built\hyphensym in ``SLS'' color
map\footnote{more details are available online at {\tt
    http$\colon$//hea-www.harvard.edu/RD/ds9/}}. This 256\hyphensym
bit ``rainbow'' color map including black and white allows our eyes to
better distinguish between different detected objects. However, when
the total number of detected objects ($N_t$) is greater than $\sim$40
even this color map is insufficient to distinguish between all unique
neighboring sources.  As a result, many unique objects may appear at
roughly the same color, although these are not necessarily detected as
the same object. This limitation of the color map does {\it not}
affect our calculation of $N_t$. Furthermore, for most galaxies the
segmentation maps show a number of objects near the edge of the
image. Although some of these edge detections may be related to real
dust features, we excluded these edge detections in the subsequent
analysis and discussion.

A comparison of the segmentation map and the galaxy core images
suggests that our general \sex~technique is remarkably successful in
recovering {\it only} those dust features that we identified first by
visual inspection.  Specifically, the dust feature recovery rate using
the IUM technique is very good for the majority ($>$95\%) of the
catalog.  For example, bar and spiral arm\hyphensym like features are
well-recovered as unique objects (see, e.g., MARK1330). The fidelity
of the object detection of the spiral arm features is often high
enough in these galaxies (see,e .g., NGC3081) that the spiral arm
features in the image are entirely reproduced in the corresponding
segmentation map.

Galaxies with relatively many dust features\longdash in both regular
or chaotic spatial distributions\longdash also appear to be faithfully
reproduced in their associated segmentation maps.  For example, the
regular structures in NGC1068 and NGC1066 are detected with \sex~as
are the more chaotic dust features, as seen in ES0137\hyphensym 634
and ESO323\hyphensym G77. An interesting result of this IUM analysis
is that the objects in some galaxies (e.g., NGC1386, NGC1672) are
sometimes limited to particular quadrants of what appears to be a disk
in the original image.  The distribution of dust features suggests to
the eye that this disk (in which the features are embedded) is
moderately inclined towards the viewer.  A discussion of molecular
toroid inclination effects are beyond the scope of this work, but we
will consider this result in future work. We note here that this disk
inclination was identified first and {\it only} by using the
\sex~technique.

In images where the stellar light profile is exceptionally smooth and
few dust features are identified by visual inspection, the IUM
technique may detect objects that do not strongly correlate with the
dust features visually identified in \S \ref{sec:visclass}.  This may
represent a limitation of our IUM technique. In Figure
\ref{fig:fourpanels}, we included the images of two galaxies (NGC1058
and NGC3608)\footnote{Only four galaxies\longdash MARK348, MARK352,
  NGC1058, NGC3608\longdash showed any strong distinction between the
  number, size, and spatial distribution of objects detected with
  \sex~segmentation map and dust features noted by visual inspection
  in \S \ref{sec:visclass}.} that represent this small fraction
($<5\%$) of the catalog galaxies.  We do not remove these galaxies
from the subsequent analysis for completeness and to illustrate to the
reader instances when our \sex~technique may be limited in its ability
to discern visually identified dust features.  We caution that object
detection in these few galaxies using our IUM technique may be more
sensitive to local pixel\hyphensym to\hyphensym pixel noise variations
than it is to signal variations arising from dust absorption.

Furthermore, in those galaxies that appear to have photo\hyphensym
ionization structures (see \S\ref{sec:visclass}), the number and
distribution of dust features using the IUM technique still appears to
be very well-correlated with the visually-identified dust features
(see, e.g., NGC3393 in Figure \ref{fig:fourpanels}). In some galaxies,
the potential photo\hyphensym ionization structure appears to be the
brightest structure visible in the image.  It does not appear that
this high contrast between photo\hyphensym ionization structures and
the visually-identified galactic dust structure has strongly affected
the dust feature detection using the IUM technique. Variations in the
mean signal across these structures could affect the calculation of
the local sky background with \sex, and thus influence dust feature
detection in those galaxies with possible photo-ionization emission
structures.  For example, such variations could explain the
segmentation of what appears as one chaotic dusty region into two
approximately equal area dust features along the outer edge of the
northeastern ``spiral-arm'' photo-ionization structure in NGC3393.

We also provided in Figure \ref{fig:fourpanels} two measurements of
the characteristics of the dust structure quantified with the IUM
technique. We plot the cumulative number of objects for each galaxy
contained within circular annuli centered at each galaxy's core for a
radius $r_c$, where $r_c=n\times\Delta r$ and $\Delta r$=2.0
pixels. Although we present square images of the galaxy, we only
calculate the cumulative number for annuli with radii less than 1 kpc
in the frame to remove edge detections. Using the cumulative object
number distribution, we calculate a half\hyphensym object radius
($r_{half}$) defined as the radius (in pixels units) of the annulus
that contains the inner 50\% of the total number of detected objects
in each galaxy. This value is provided in physical units (parsecs)
with measurement uncertainties in Table \ref{tab:fitpars}.

In Figure \ref{fig:alphadist}(a) we plot the distributions of
$r_{half}$.  We fit a Gaussian function to the distribution of
$r_{half}$ for \syall~galaxies and provide the parameters of the
best\hyphensym fit functions in Table \ref{tab:spartab2}.  There is no
apparent distinction in the distribution of half\hyphensym object
radii between \syall~galaxies. This is confirmed by a two\hyphensym
sample K\hyphensym S test, the results of which indicate that the
parent distributions from which the half\hyphensym object radii
distribution were drawn are not likely to be unique.

Figure \ref{fig:fourpanels} also included the object surface density
distribution ($\Sigma$) measured for the galaxies, which we defined as
$\colon$
\begin{equation}
\Sigma = \mbox{log}\left(\frac{N}{4\pi{({\mbox r}_2-{\mbox
 r}_1})^2}\right)
\end{equation}
where N is the number of objects contained within annuli of width
equal to 10 pixels. We fit a linear function to the object surface
density function versus radius and measure the best\hyphensym fit
slope ($\alpha$). In Figure \ref{fig:alphadist}b, we provide the
distribution of $\alpha$ measured. We fit a Gaussian function to the
distribution of $\alpha$ as measured for the two classes of AGN, and
measure the Gaussian centroids and FWHM to be nearly equal (Table
\ref{tab:spartab2}). The similarity in the distributions is confirmed
by a two-sample K\hyphensym S test. Thus, the distribution of the
object surface density functions does not appear to be unique to the
class of AGN.

In Figure \ref{fig:coverfrac}a, we plot the number of objects ($N_t$)
identified in each galaxy.  We measure the centroid and half\hyphensym
width half maximum (HWHM) of both the \syall~distributions (see Table
\ref{tab:spartab2}). We fit a Lorentzian function, rather than a
Gaussian function, to better account for the broad extension from the
HWHM peak to large object numbers in the $N_t$ distribution.  The
centroid of the best\hyphensym fit Lorentzian function of objects
equals to $\sim$18 for both classes of AGN, but the mean value of the
\syall~distributions equals to 47 and 35, respectively. Thus, these
distributions appear to be significantly different. We confirm this
result via a two\hyphensym sample K\hyphensym S test$\colon$ we
measure {\it d=}0.35 and {\it p}=0.01, and conclude the empirical
distributions of $N_t$ measured for the \syall~populations are likely
drawn from independent parent distributions. If the objects detected
with \sex~physically correspond to dust features in the galaxies, then
we may conclude that Sy2 galaxies are, on average, dustier than Sy1
galaxies. If we remove the four galaxies discussed above for which our
\sex~technique did not appear to detect objects that are closely
associated with the dust features that we identified by visual
inspection, though, we measure the K\hyphensym S test probability
statistic for the distributions of $N_t$ equal to {\it p}=0.06.  In
this case, we can not reject the null hypothesis, and instead are
forced to conclude that the distributions of $N_t$ measured for
\syall, respectively, were likely drawn from the same parent
population.

The \malkrel$\,$ did not consider the number of dust features
explicitly, but the assignment of relative degrees of dustiness to
galaxies implicitly reflects the number of dust features that were
identified visually. It seems to the authors that is easier to
visually classify the dust structure as irregular if it contains a
relatively large number of dust features, because the eye can more
readily identify absorption patterns and divergences therefrom.  Thus,
the mean $N_t$ measured for \syall~ (Figure \ref{fig:coverfrac}a) may
support\longdash albeit indirectly\longdash the \malkrel$\,$ in a
quantitative way.

In Figure \ref{fig:coverfrac}b and \ref{fig:coverfrac}c, we also
provide the covering fraction ($f_c$) and the average number of pixels
($N_p$) associated with objects detected by \sex. We fit a Gaussian
function to the distributions measured for each of these parameters,
and observe no distinction between the Gaussian centroid or FWHM
measured for \syall, respectively (see Table \ref{tab:spartab2}). We
confirm the similarity between the measured distributions by a
K\hyphensym S test, and conclude that these distributions are likely
drawn from the same parent distribution. These results would not
support the \malkrel, or at least not demand it.

Throughout this work we have considered the results of our {\it only}
in the context of the Unified Model, as outlined in \cite{A93}.  We
restricted our discussion of these results to this context, in part,
because we were motivated in this work to extend the analysis first
presented in \cite{M98}, in which the authors make a similar
assumption on the nature of AGN.  The assumption of this model is
still fair; despite extensive debate (see \S\ref{sec:intro}) the Model
provides a remarkably robust explanation for the observed diversity of
AGN\footnote{If only because many of the systematic considerations of
  the Unified Model are still, regrettably, limited by large measured
  uncertainties; cf. Guainazzi et al. 2011}.  But this model is not
without rivals. For example, the ``clumpy torus'' model reduces the
thick, dusty torus---the inclination of which gives rise to the
observed dichotomy of Seyfert-type AGN---to distinct individual dust
clumps that are generally distributed about the central engine.  In
this model, the AGN type that one observes is not a ``binary''
function of perspective; rather, the probability of observing a Type 1
AGN decreases as the viewer moves towards an ``edge-on'' perspective
but never reaches zero.  We observe a core region that is hundreds of
parsecs beyond the toroid, though.  Thus, a full interpretation of our
results in the context of this model is beyond the scope of this
project and we reserve that discussion for future work.

\section{Conclusions and Summary}\label{sec:conclusion}

Recently, multi\hyphensym wavelength high spatial\hyphensym resolution
images and spectroscopy of AGN\hyphensym host galaxies have revealed
that the characteristics (e.g., morphology, dust, properties of the
emission line regions; see \S \ref{sec:intro}) of these galaxies may
not be consistent with the Unified Model of AGN. In particular, {\it HST}
WFPC2 imaging of the cores of local ($z$\,\,\lsim\,\,0.035)
Seyfert\hyphensym host galaxies established that the morphologies of
dust features in these galaxies correlates with the AGN class
(\malkcit). We investigated this trend by visually inspecting a
catalog of archival WFPC2 F606W images of the cores of 85 local
($z<0.015$) Seyfert galaxies, and classified these AGN by the presence
and distribution of dust features in each. We determined that Sy2
galaxies were more likely to be associated with galaxies whose core
dust morphology is {\it more} irregular and of ``later\hyphensym
type'' morphology. Thus, our visual classification of our Seyfert
galaxies confirms the qualitative morphological relationship
established by \malkcit. We concur with the conclusion of
\malkcit~that\longdash if this morphological relationship is
indicative of a fundamental distinction between the subclasses of
AGN\longdash this result weakens the central postulate of the Unified
Model of AGN.

We extended the study of this morphological relationship established
through the qualitative visual method by re\hyphensym\,analyzing the
images using quantitative morphological tools. First, in \S
\ref{subsec:methodol} and \S \ref{subsec:origanaly}, we developed and
measured the $C^*,A^*,S^*$ and $G^*$\hyphensym$M^*_{20}$ parameters
for the galaxies. The distribution of these parameters as measured for
\syall~AGN did not strongly distinguish between the Seyfert class and
morphology of the host galaxy. We determined that the parameter
distributions for \syall~AGN are likely drawn from the {\it same}
parent distribution using a two\hyphensym sample K\hyphensym S test,
with the exception of the concentration $C^*$ parameter. In comparison
with the other four parameters in this set, though, $C^*$ is the least
sensitive parameter to the morphological features identified and
classified by visual inspection. We conclude from this analysis that
no strong morphological distinction exist between the cores of the
\syall~AGN host galaxies. This conclusion conflicts with the
established \malkrel$\,$ which we have visually confirmed in \S
\ref{sec:visclass}. 

In order to resolve these apparently conflicting conclusions and to
address the possibility that the $C^*,A^*,S^*$ and $G^*$\hyphensym$M^*_{20}$ are
less effective tools for characterizing sub\hyphensym kiloparsec scale
dust features, we developed a new method to quantify the core dust
morphologies of the AGN galaxies.  This method combines \sex~with the
IUM technique. We applied this method and found that the distributions
of the average number of detected dust features in \syall~AGN may be
different.  Thus, we have measured a quantitative distinction between
\syall~AGN that supports the \malkrel. Yet, there was no concordance
between this result and any other result (i.e., the radial
distribution, size and covering fraction of dust features) derived
from this quantitative method. We therefore cannot strongly
distinguish between \syall~AGN on the basis of their core morphologies
using this quantitative method.

In conclusion, we studied the relation between the host galaxy core
($r<1$ kpc) morphology and the Seyfert class of AGN. We find evidence
to support this conclusion from the results of a quantitative
assessment of the core dust morphology using existing and new
methods. However, we can not strongly distinguish between the core
dust morphology of \syall~AGN using the complete set of morphological
parameters. Thus, we are reluctant to suggest that the Unified Model
of AGN must be significantly modified to accommodate the results of
this qualitative and quantitative analysis. In the future, better and
more internally consistent {\it qualitative} and {\it quantitative}
methods need to be developed to elucidate the true nature of the cores
of Seyfert galaxies. We expect that JWST will be able to do a similar
investigation of this topic at longer wavelengths, thereby penetrating
deeper into the Seyfert galaxies central dust.

\section{Acknowledgements}

 This work is supported by NASA ADAP grant NNX10AD77G (PI: Windhorst).
 The Hubble Space Telescope is operated by the STScI, which is
 operated by the Association of Universities for Research Inc., under
 NASA contract NAS 5\hyphensym 26555. We make extensive use of the
 HLA, the online repository of {\it HST} data that is maintained as a
 service for the community by the Space Telescope Science Institute,
 and SDSS Data Release 7. The authors acknowledge S. Kenyon,
 R. Jansen, S. Cohen, J. Bruursema, M. Benton and Y. K. Sheen for
 helpful technical support and scientific discussion. M. Rutkowski
 acknowledges the U.S. State Department Fulbright Program for
 financial support of this research.

\appendix
\renewcommand\section{\begin{centering}{\normalfont\bfseries Appendix }\end{centering}}

\section{\bf{A. Spatial Resolution$\colon$ Ground vs. Space\hyphensym based imaging}}\label{sec:SpatScale}

We implicitly assumed throughout this paper that {\it HST} images are
necessary to conduct the quantitative morphological analyses. If
lower spatial resolution ground\hyphensym based optical images could
be used instead of the high spatial resolution {\it HST} images, we could
significantly increase the number of galaxies we can consider. The SDSS
archive, for example, could provide images of hundreds of local AGN.

We downloaded SDSS r$^{\prime}$ images for 7 AGN that were in both the
SDSS DR7 archive and our catalog presented in \S
\ref{sec:thedataset}. We made thumbnails of the core ($r<$1 kpc) SDSS
images of each galaxies and measured $C^*,A^*,$ and $S^*$ parameters
using the same techniques outlined in \S \ref{subsec:methodol} for
each galaxy. In Figure \ref{fig:appendixa}, we compare these
measurements with those presented in \S\ref{sec:modginim20} which were
measured in {\it HST} F606W images. It is apparent from this comparison that
$A^*$ and $S^*$ cannot effectively discriminate between the
morphologies of the SDSS galaxies. This result confirms that the
quantitative morphological analysis we performed above requires the
high spatial resolution {\it HST} images.

\section{{\bf B. Size\hyphensym Scale Relation}}\label{sec:SizeScale}

Two galaxies that are identical in every possible way (e.g.,
morphologically), but at significantly different distances from an
observer, will appear different in images obtained with the same
telescope, because each CCD pixel covers an intrinsically larger
physical area in the more distant galaxy. As a result, the dust
features in the more distant galaxy are less well\hyphensym resolved
spatially. Our catalog includes galaxies in the range between 0.001
$<z<$ 0.015, or equivalently a factor of 10\hyphensym 15 in physical
distance.

We selected six galaxies\longdash NGC1068, NGC3185, NGC3227, NGC3608,
NGC4725, NGC4941\longdash with morphologies that represented the
diversity of morphologies in our catalog. These galaxies are all at
distances $\simeq$ 15 Mpc, and we use these galaxies to quantify the
extent to which we are able to identify or measure dust features in our
catalog galaxies as a function of distance. We do not use the nearest
galaxies ($D$\,\,\lsim\,\,10Mpc) in our catalog because these galaxies
include large off\hyphensym chip regions that significantly
affect the measurement of asymmetry.

We then rebinned each of these galaxies to a pixel scale, {\it s},
such that $\colon$
\begin{equation}
s = \ell_{GAL} \times \frac{D_{GAL}}{D_{z=0.015}},
\end{equation}
where $\ell$ is number of WFPC2 0\farcs10 pixels spanning 1000 pc at
the physical distance, {\it D}, to the galaxy and $D_{z=0.015}$
corresponds to the distance to a galaxy at the upper redshift
range of galaxies in our catalog (63 Mpc).

We measure $C^*,A^*,S^*$ and $G^*$\hyphensym$M^*_{20}$ for these
artificially\hyphensym redshifted galaxies and compare the measured
values with the original measurements (\S
\ref{subsec:origanaly}). This comparison is presented in Table
\ref{tab:appendixb} as $\delta=\frac{|X-Y|}{Y}$, where {\it X} and
{\it Y} are the morphological parameters measured in galaxy images
at $D_{gal}$ and artificially redshifted to $D_{z=0.015}$.

In general, the measurement of these parameters does not seem to be
strongly affected by the relative distance of the galaxy, at least
over the relatively small redshift range that we consider in this
project. For all parameters, $\delta$ is much smaller than the
measured dispersion in the range of parameters measured in \S
\ref{subsec:origanaly}. We conclude that range of measured parameters
(see Figures \ref{fig:CGM20plot} and \ref{fig:asymdist}) are
indicative of {\it morphological} distinctions between the cores of
the sample galaxies, as we assumed in the discussion in
\S\ref{subsec:origanaly}.

\section{{\bf C. Sensitivity of measurements to the estimated sky\hyphensym background}}\label{sec:appendixc}
 
Windhorst et al.~(in prep.) measured the surface brightness of the
zodiacal background as function of $\ell^{Eq.}$ \& $b^{Eq.}$ from 6600
WFPC2 F606W and F814W Archival dark\hyphensym time images. We
reproduce these measurements for the F606W zodiacal background from
Windhorst et al.~(in prep.) in Figure \ref{fig:zodiplot}. In
\S\ref{subsec:methodol} we estimated the surface brightness of the
zodiacal background along the line\hyphensym of\hyphensym sight to
each galaxy in our catalog. We made the reasonable assumption that the
only background emission present in the images in the cores of our
catalog galaxies arises from the zodiacal background and then
corrected for this background alone in each image.

In this section, we measure the uncertainty in our measurements of
$C^*,A^*,S^*$ and $G^*$\hyphensym$M^*_{20}$ and the object surface
density distribution associated with our assumption of the brightness
of the background in the images. In general, the brightness of the
background is determined to have a minimal effect on these parameters,
with the notable exception of $G^*$, and to a lesser extent, the
$\alpha$ slope parameters. In Figure 10, we compare the measurements
of $G^*$ for galaxies corrected for a zodiacal background equal
to$\colon$ a) zero ($G^*_a$); b) the Windhorst et al.~background
($G^*_b$); and c) a hypothetical background 10 times {\it larger} than
the measured in Windhorst et al. The latter estimate of the background
emission is highly unlikely in any {\it HST} image (see Figure
\ref{fig:zodiplot}). We assume such a large background here only to
provide an upper extremum to our measurement of the effect of the
background surface brightness assumption.

The dispersion measured for most parameters, i.e., $C^*,A^*,S^*$ and
$M^*_{20}$, for different estimates of the zodiacal surface brightness
was small ($<$1\%).  There is a large dispersion between $G^*_a$ and
$G^*_c$.  We attribute this dispersion to the removal of relatively
faint pixels from the measurement of $G^*$ as increasingly larger
values for the sky surface brightnesses are subtracted from the
images.  This has the net effect of (artificially) enhancing the flux
associated with relatively higher signal pixels, which increases
$G^*$.

We note that there is a modest increase in the measurement of $\alpha$
(the slope of the object surface density function), when comparing
cases (a) and (c). The net effect of background on this measurement is
less than 5\%.  Hence, blindly adopting the most likely zodiacal
sky\hyphensym brightness as a function of $\ell^{Eq.}$ \&
$b^{Eq.}$\longdash when this background is not directly
measurable\longdash is an acceptable and, in our case the only viable,
approach.

\section{\bf{D. IUM Technique$\colon$ Dust Feature Detection Threshold }}\label{sec:deteff}

The detection of dust features with Source Extractor is explicitly
dependent on the detection parameters defined by the user in the
configuration file. Here we discuss the typical contrast level of the
dust features, relative to the ``sky background'' in the images, which
\sex~detected for those parameters outlined in
\S\ref{subsec:sextractmeth}.  We define the ``contrast'' as:

\begin{equation}
  \mbox{Contrast}=\frac{f_{dust}-f_{meansky}}{f_{dust} + f_{meansky}}\times 100\%, 
\end{equation}
where $f_{dust}$ is the flux associated with a detected object using
the IUM technique and $f_{meansky}$ is the average sky value measured
in a uniform ``sky'' region drawn from the core image.

We measured the contrast parameters for two representative galaxies in
the sample, NGC3081 and NGC3608. The IUM technique appears to work
very well in detecting the dust clumps in NGC3081, whereas NGC3608 was
largely devoid of dust clumps according to the visual inspection.  For
each of these galaxies, we measured the contrast values for three
detected dust clumps, using two relatively large but smooth ``sky''
regions (Area$\simeq$100---200 sq. pixels).  The mean contrast,
($f_{dust}=\bar{f}$, the average flux associated with the dust
feature) measured for NGC3081 and NGC3608 equals 6 and 2\%,
respectively.  Assuming $f_{dust}$ equal to the flux of the {\it
  brightest} pixel in each of the dust features, the mean contrast is
measured to 12\% and 4\% for the two galaxies. We measure the relative
height of the mean flux associated with the dust features above the
mean sky equal to 50--90$\times\sigma_{sky}$ for NGC3608 and NGC3081,
respectively.  We note here the fainter sources could be detected if
the \sex~detection parameters are revised, but this could introduce
more ``false positive'' dust feature detections and would fragment
coherent visible structure.

\setlength{\abovecaptionskip}{0pt}
\setlength{\belowcaptionskip}{0pt}
      
\long\def\@makecaption#1#2{%
\vskip\abovecaptionskip
\sbox\@tempboxa{#1: #2}%
\ifdim \wd\@tempboxa >\hsize
#1: #2\par
\else
\global \@minipagefalse
\hb@xt@\hsize{\box\@tempboxa\hfil}%
\fi
\vskip\belowcaptionskip}
\makeatother
 
\begin{figure}
\centering
\begin{tabular}{c}
\includegraphics[width=6in,angle=0,scale=1]{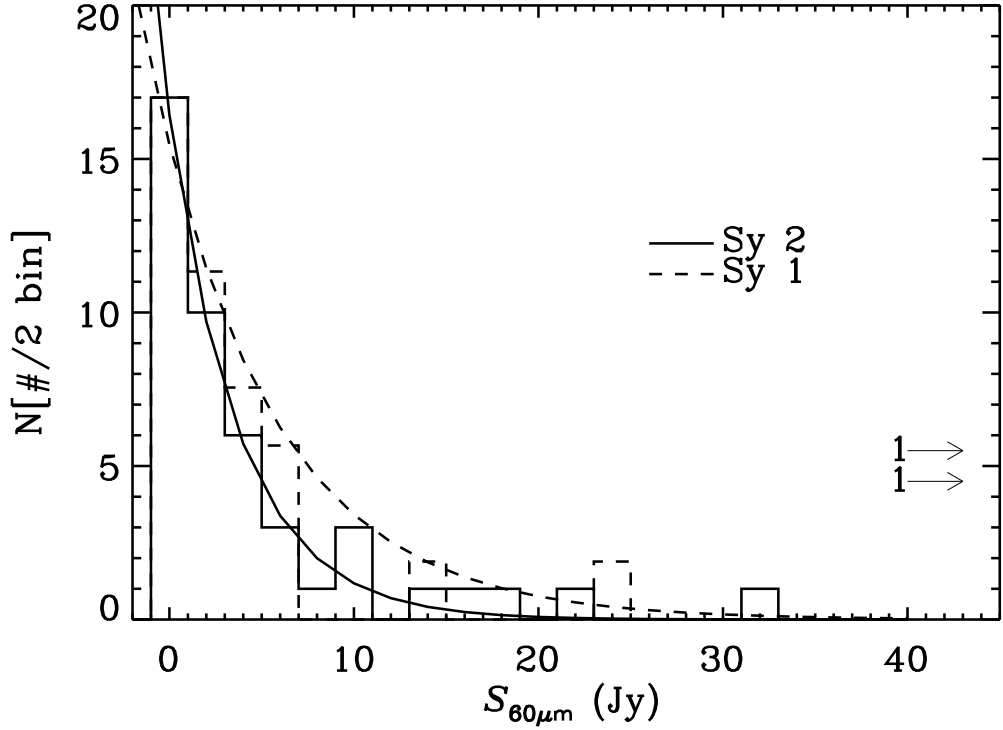} \\
\end{tabular}
\caption{The FIR flux distribution of the catalog Seyfert galaxies from the IRAS Faint Source catalog (available via NED).  The Sy1 distribution has been scaled to match the peak in the Sy2 distribution and both samples each had one AGN with measured FIR flux greater than 40 Jy (illustrated by arrows).  We fitted an exponential function ($\propto exp[-f/\tau])$, where $\tau$=3.8 \& 6.7 for Sy1 and Sy2, respectively.  We did not select Seyfert AGN on the basis of their FIR properties, but the samples appear to be generally similar, with the caveat that the sample has a known bias towards Sy2 AGN.}
\label{fig:oiifir}
\end{figure} 
\newpage
\clearpage
\addtocounter{figure}{-2}
\setlength{\abovecaptionskip}{-15pt}
\setlength{\belowcaptionskip}{-10pt}
\long\def\@makecaption#1#2{%
\vskip\abovecaptionskip
\sbox\@tempboxa{#1: #2}%
\ifdim \wd\@tempboxa >\hsize
#1: #2\par
\else
\global \@minipagefalse
\hb@xt@\hsize{\box\@tempboxa\hfil}%
\fi
\vskip\belowcaptionskip}
\makeatother
\begin{sidewaysfigure}
\begin{minipage}[h]{1.00\textwidth}
\noindent\centerline{
\includegraphics[width=0.25\textwidth]{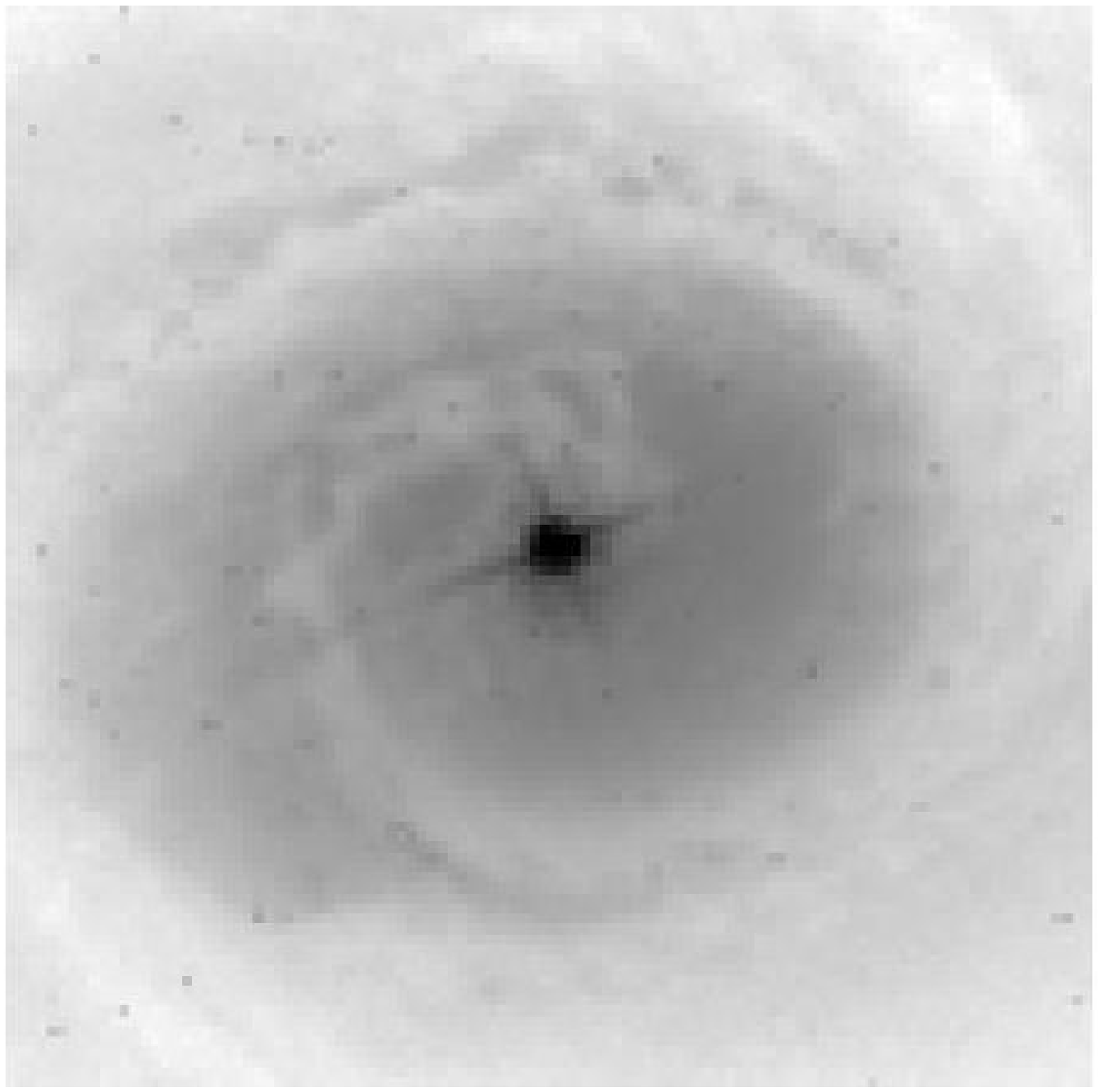} \hspace{-10mm}
\includegraphics[width=0.25\textwidth]{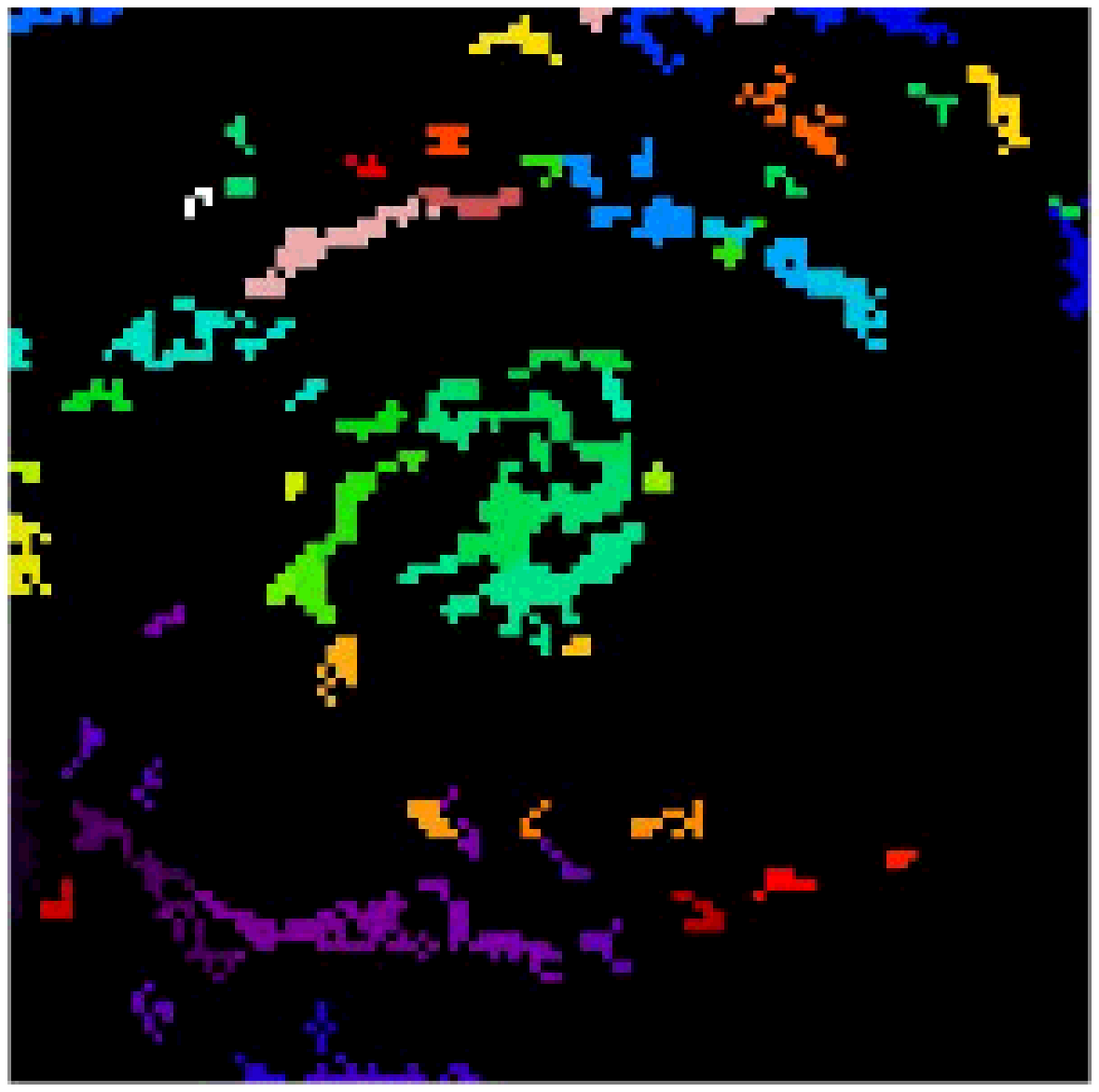}
\includegraphics[width=0.25\textwidth]{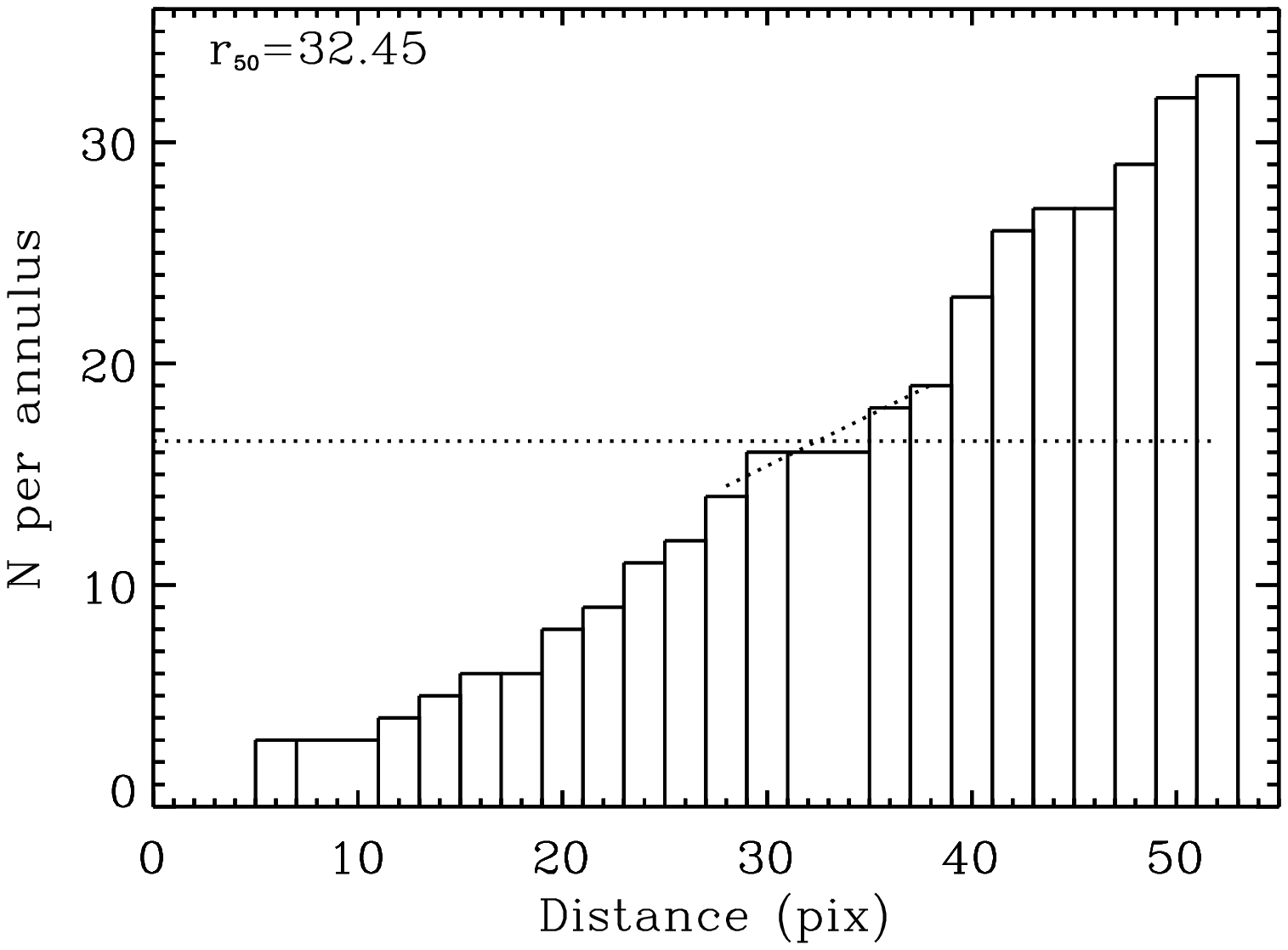}
\includegraphics[width=0.25\textwidth]{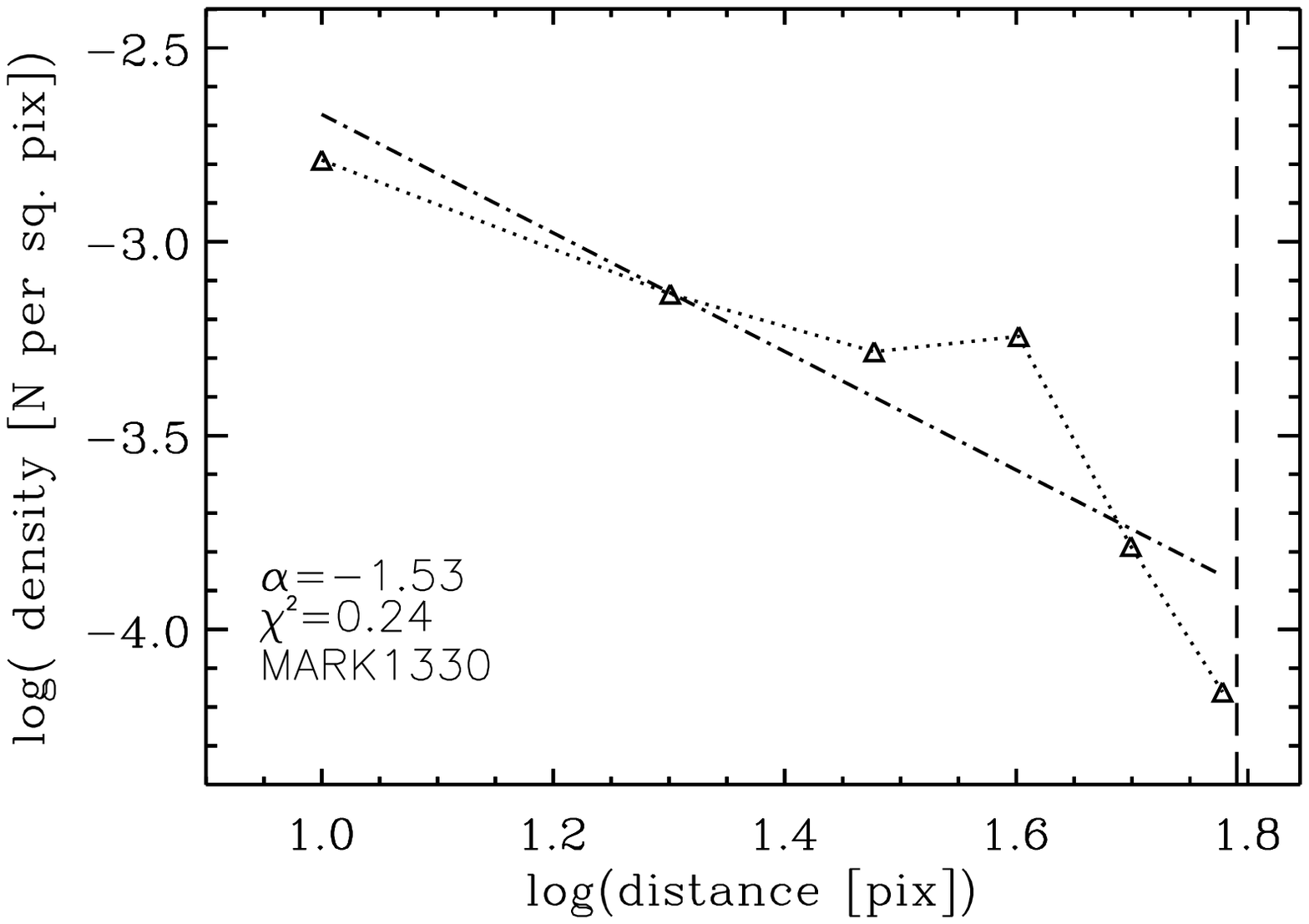}
}\\
\caption*{MARK1330}
\noindent\
\noindent\centerline{
\includegraphics[width=0.25\textwidth]{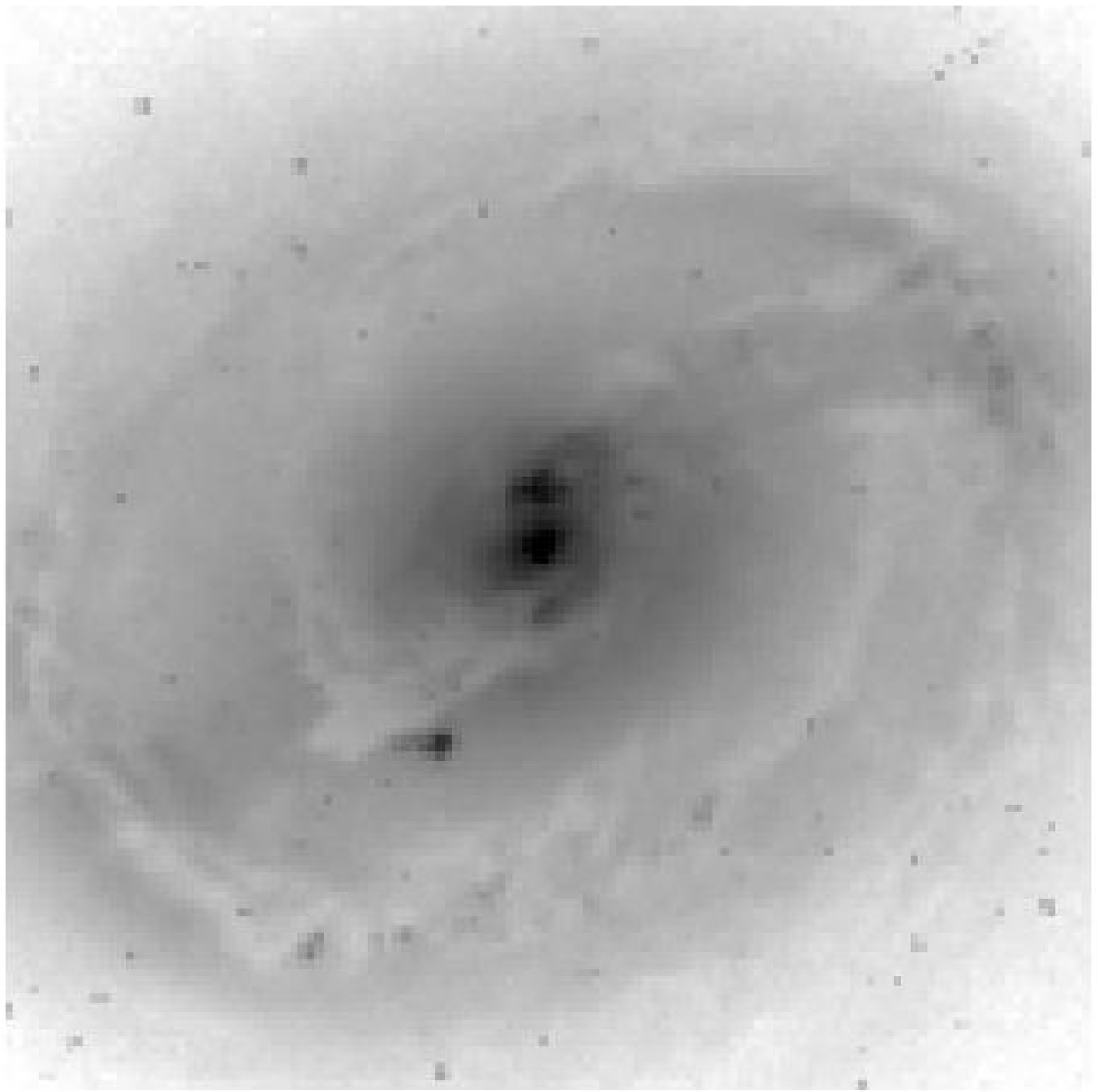} \hspace{-10mm}
\includegraphics[width=0.25\textwidth]{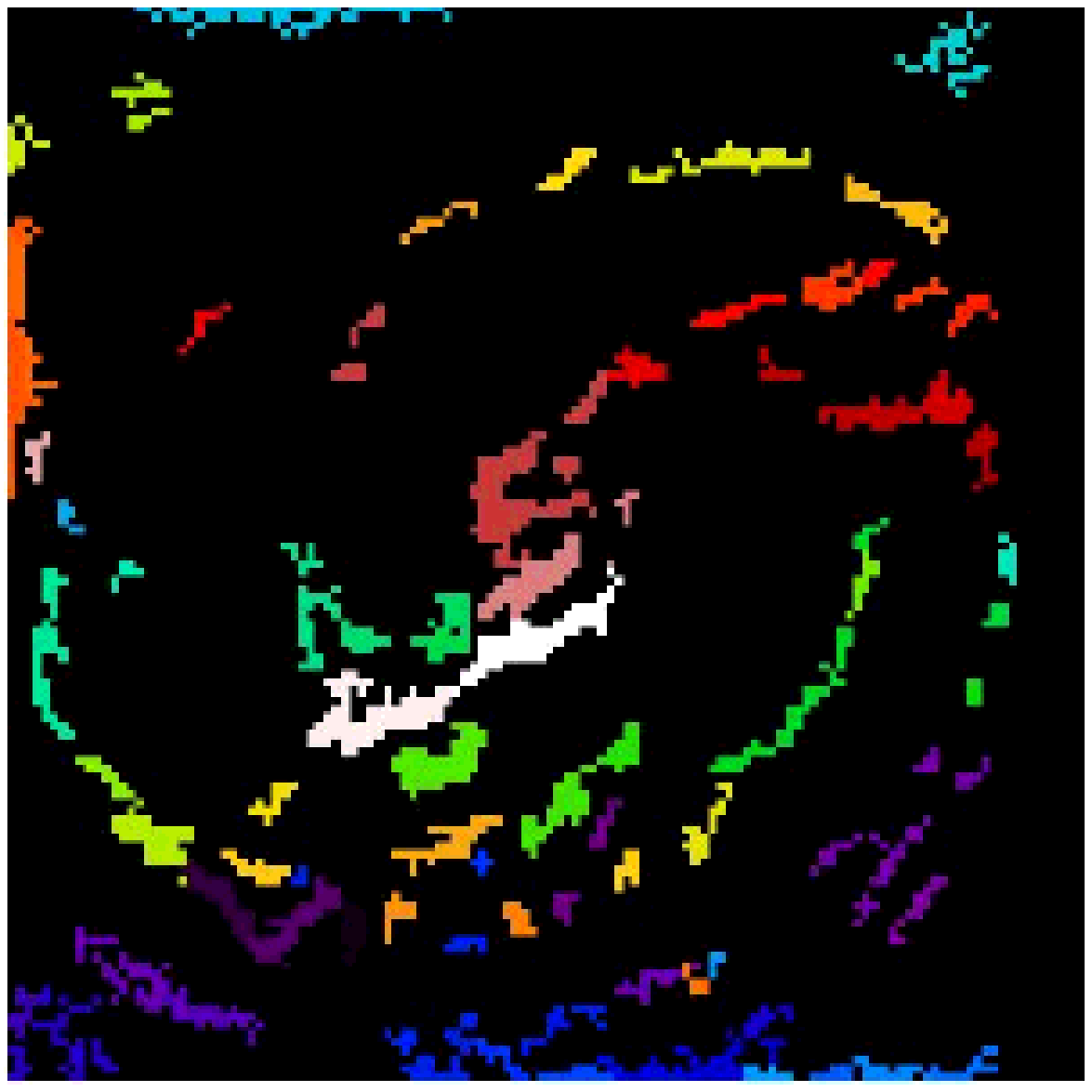}
\includegraphics[width=0.25\textwidth]{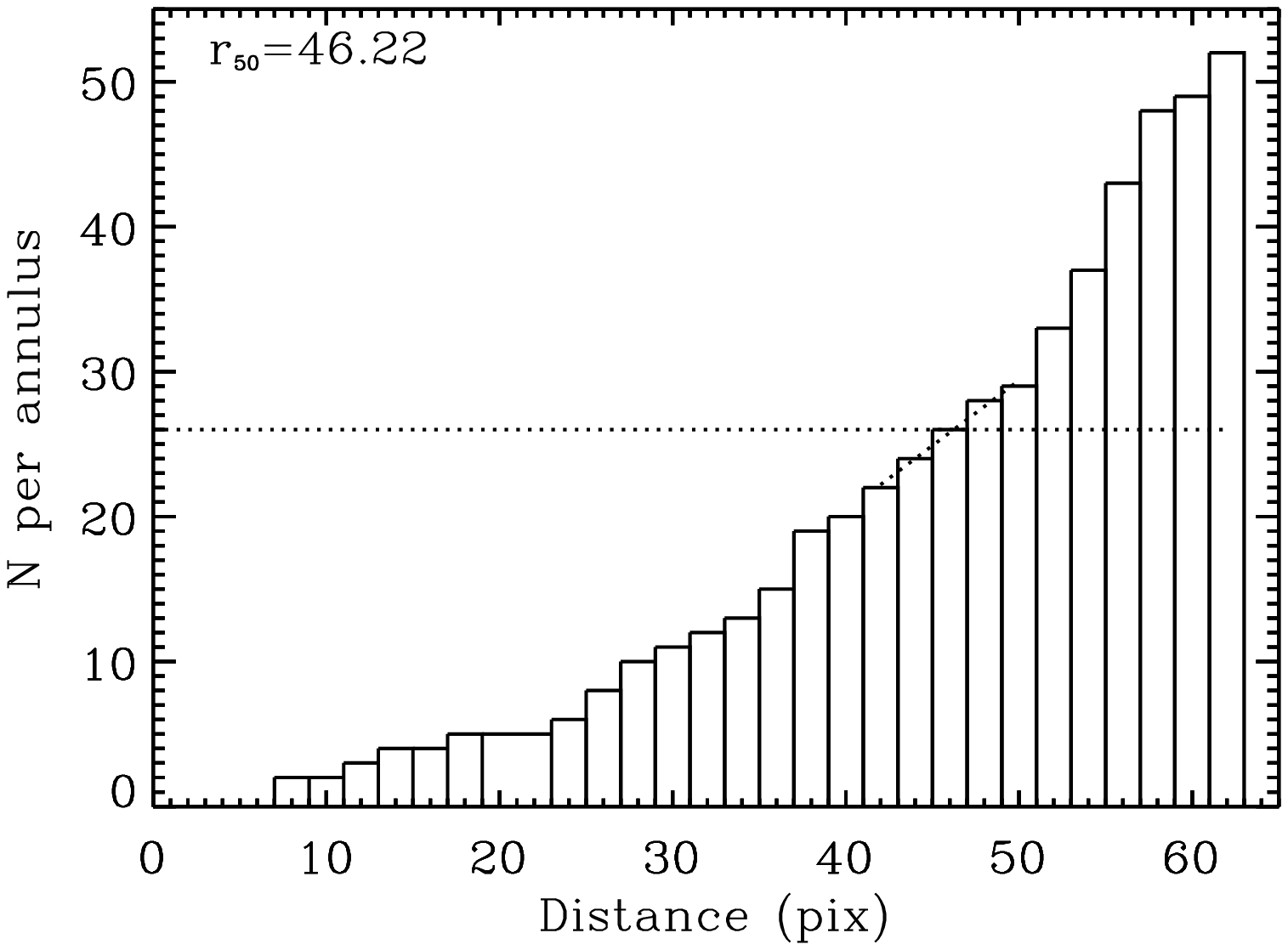}
\includegraphics[width=0.25\textwidth]{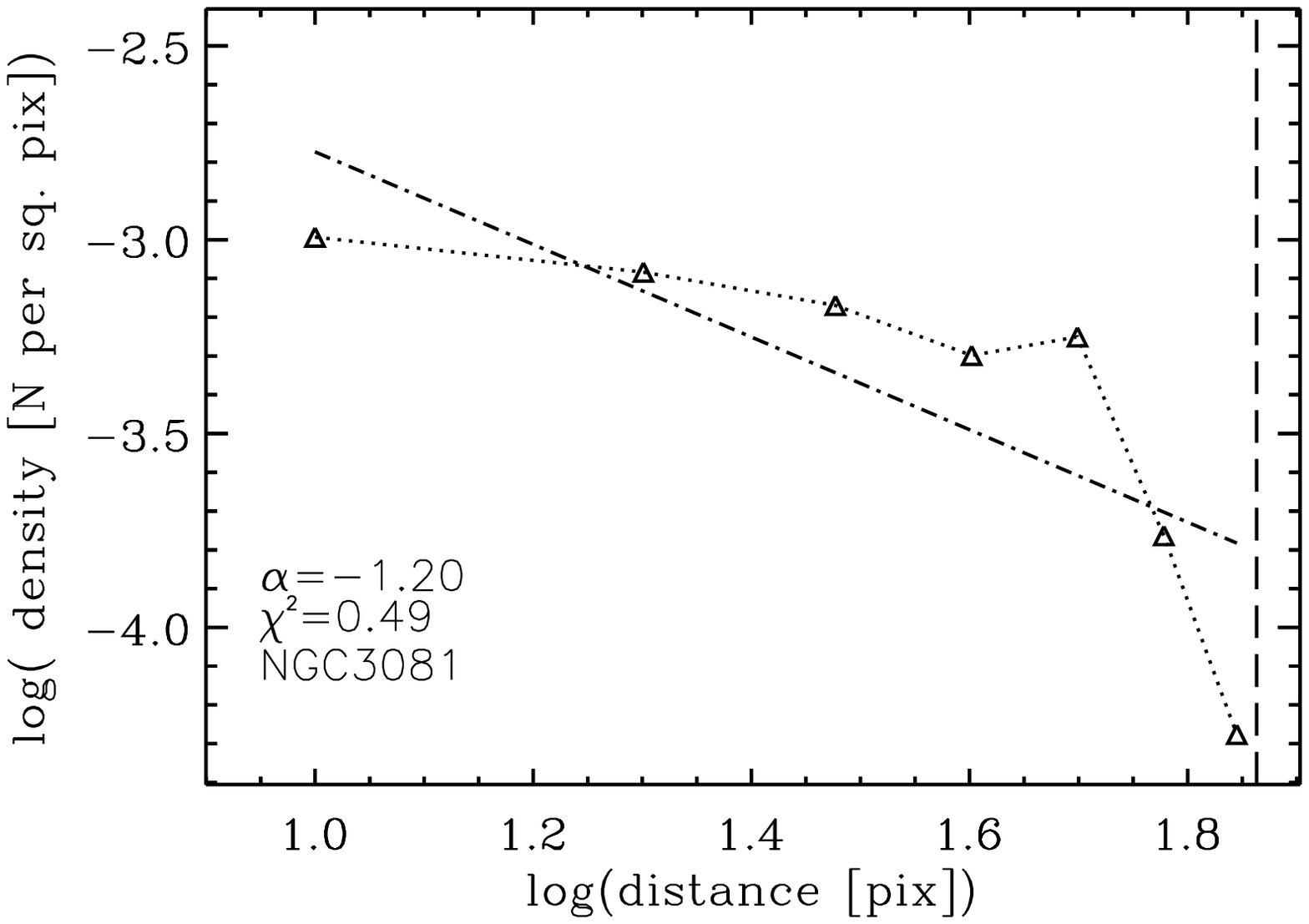}
}\\
\caption*{NGC3081}
\end{minipage}
\vspace*{+10mm}
\centering
\caption{From left to right, we provide the WFPC2 F606W postage\hyphensym stamp image of each of the catalog that was used to classify galaxy morphology qualititately (\S \ref{sec:visclass}) and quantitatively (\S\ref{sec:modginim20} and \S\ref{sec:sexintro}). Next, we provide the segmentations maps that were generated using the inverse unsharp\hyphensym mask method defined in \S \ref{subsec:sextractmeth} are provided.  Finally, we provide the cumulative number function of objects measured for radii less than 1 kpc and the half\hyphensym object radius as well as the object surface density, defined as the number of objects per annulus and the best\hyphensym fit slope $\alpha$. We discuss each of these data products at length in \S \ref{subsec:sexresults}.}
\label{fig:fourpanels}
\addtocounter{figure}{1}
\end{sidewaysfigure}
\addtocounter{figure}{-1}
\begin{figure}
\centering
\begin{tabular}{c}
\includegraphics[width=6in,angle=0,scale=1.0]{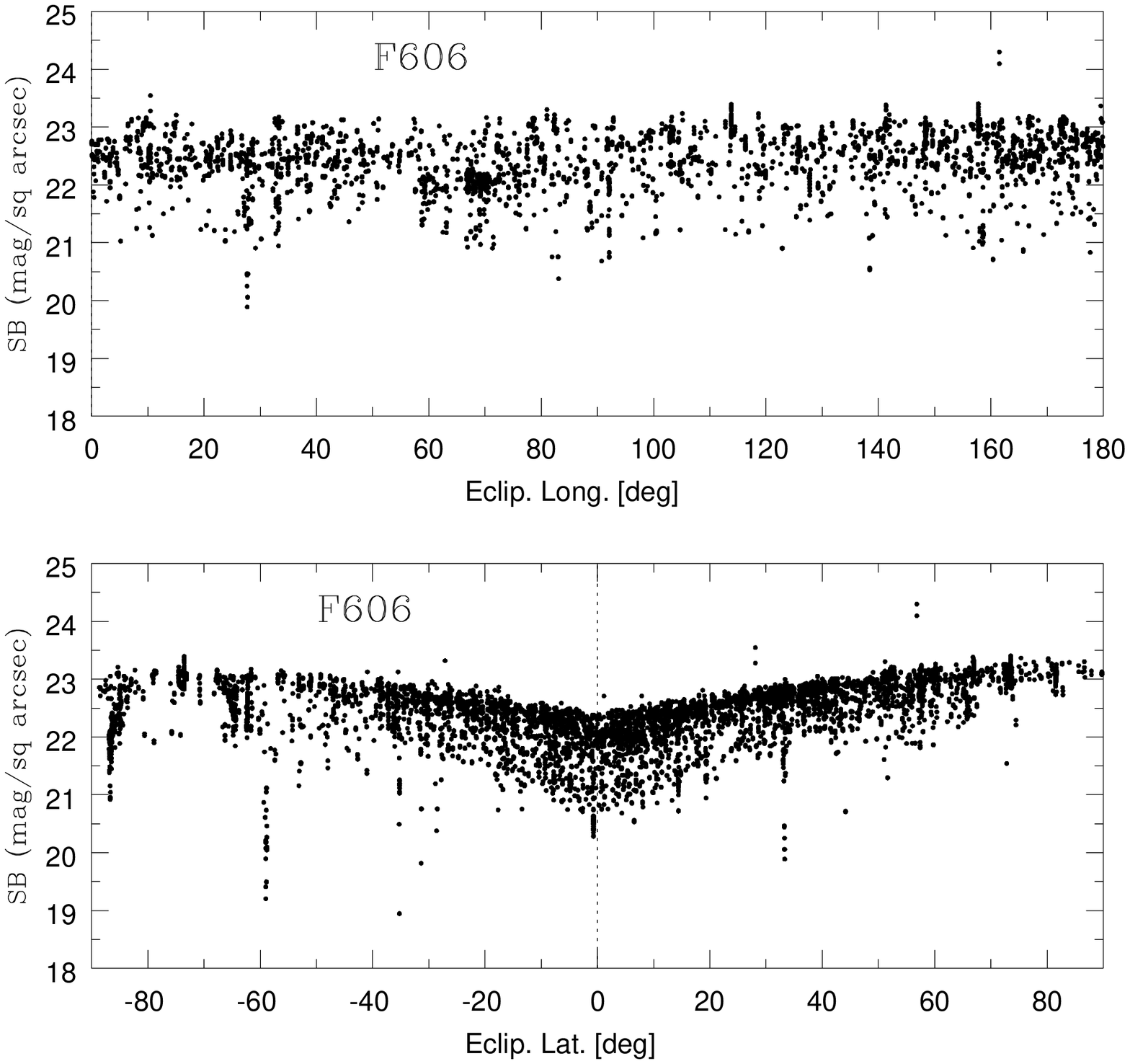} \\
\end{tabular}
\caption{Windhorst et al. (in prep.) measured the surface brightness
  of the zodiacal background as a function of ecliptic latitude and
  longitude using $\sim$6600 dark\hyphensym orbit, archival F606W
  WFPC2 images.  We estimate the surface brightness of the zodiacal
  background along the line\hyphensym of\hyphensym sight to our
  catalog galaxies, and correct for this zodiacal emission by
  subtracting the background from the core image in
  \S\ref{subsec:methodol}.}
\label{fig:zodiplot}
\end{figure}
\setlength{\abovecaptionskip}{-5pt}
\setlength{\belowcaptionskip}{-5pt}
\long\def\@makecaption#1#2{%
\vskip\abovecaptionskip
\sbox\@tempboxa{#1: #2}%
\ifdim \wd\@tempboxa >\hsize
#1: #2\par
\else
\global \@minipagefalse
\hb@xt@\hsize{\box\@tempboxa\hfil}%
\fi
\vskip\belowcaptionskip}
\makeatother
\begin{figure}
\centering
\begin{tabular}{c}
\includegraphics[width=8in,angle=0,scale=0.85]{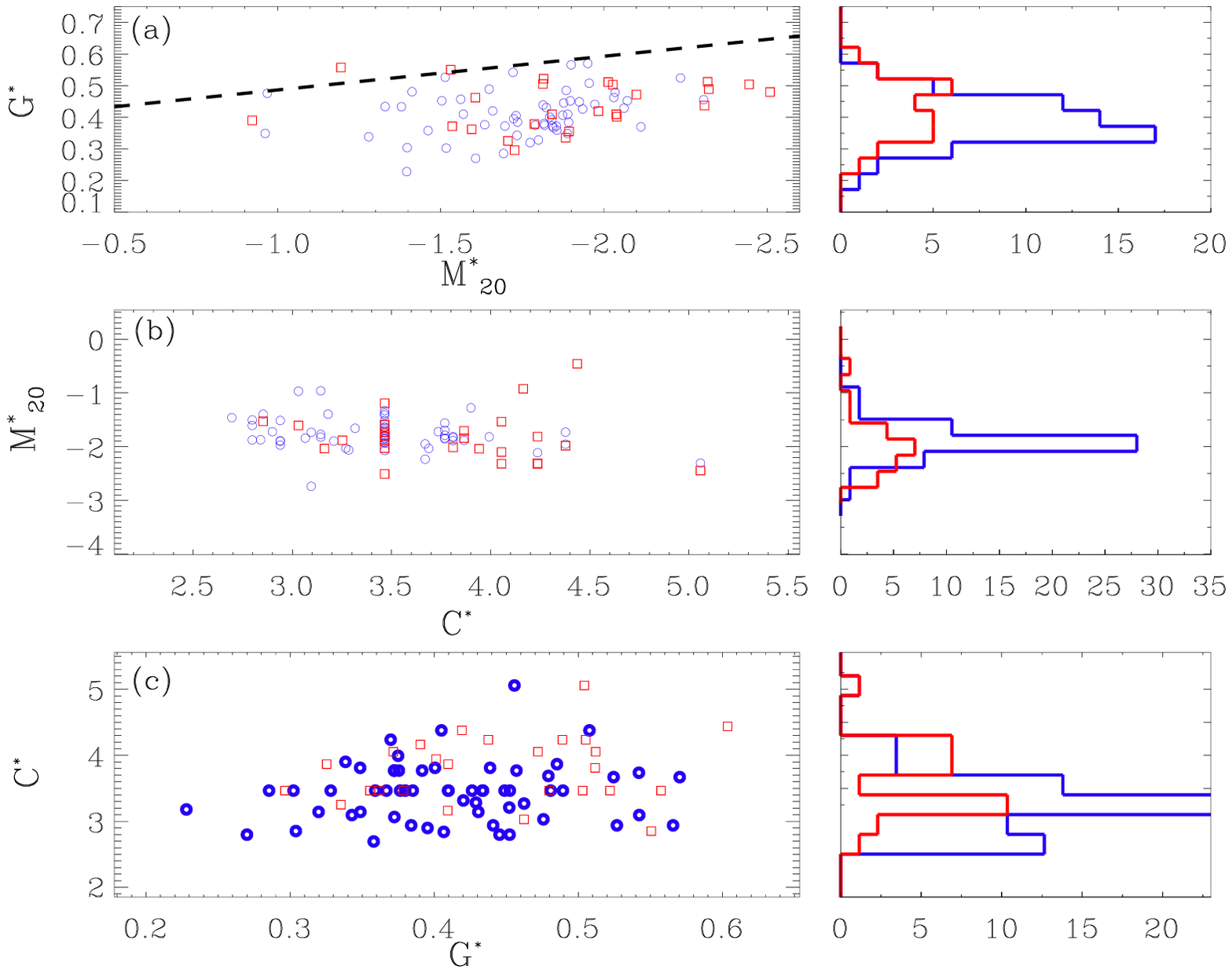} \\
\end{tabular}
\caption{The   $C^*,G^*,$   \&   $M^*_{20}$  parameters were   defined   in   \S
  \ref{sec:modginim20}.  Here, we plot the measured parameters
  for \syall~AGN as blue circles and red squares, respectively. In
  (a), we overplot the empirically\hyphensym defined line (dashed)
  that was defined by \cite{L04} to separate ``normal'' galaxies from
  Ultraluminous Infrared Galaxies (ULIRGs), but find that this line
  does not strongly differentiate starburst\hyphensym type galaxies
  from ``normal'' galaxies.  The distributions of each of these
  parameters appear indistinguishable
  for \syall~(see \S \ref{subsec:origanaly} for more details).  We
  confirm this with a two\hyphensym sample Kolmogorov\hyphensym
  Smirnov test.  The measured K\hyphensym S parameter is large for
  both the $G^*$ and $M_{20}^*$ distribution ({\it d}=0.28 and 0.29,
  respectively), but the associated probabilities are also both large
  ({\it p}=0.09 and 0.08).  Therefore, we do not reject the null
  hypothesis that both distributions are drawn from unique parent
  distribution.  However, the K\hyphensym S test for the distribution
  of $C^*$ does suggest that the measured distributions for \syall~AGN
  are drawn from unique parent distributions ({\it d}=0.38 and {\it
  p}=0.01).  We discuss these results in \S \ref{sec:conclusion}.}
\label{fig:CGM20plot}
\end{figure} 

\begin{figure}
\centering
\begin{tabular}{c}
\includegraphics[width=6in,angle=0,scale=1.0]{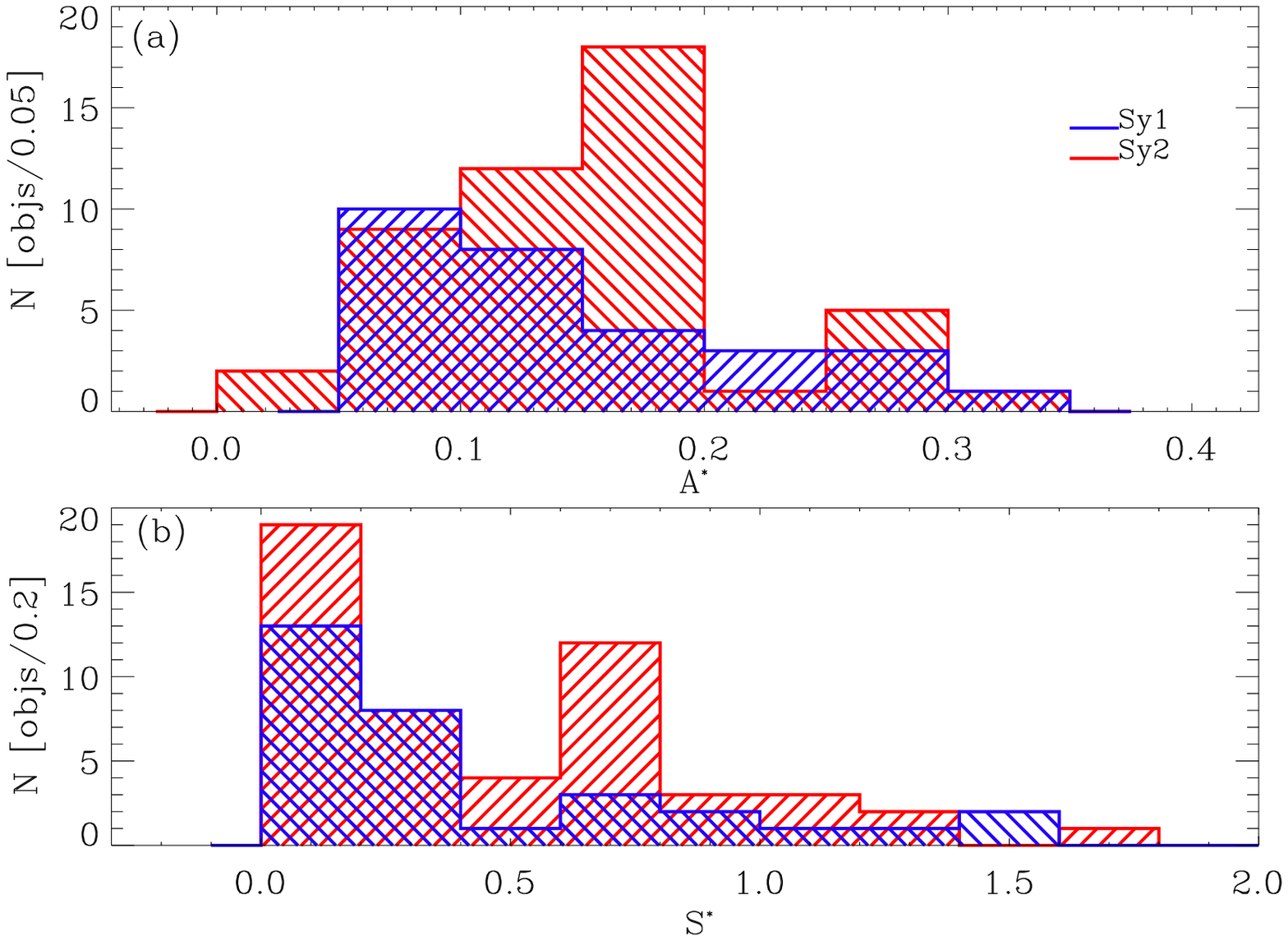} \\
\end{tabular}
\caption{The distribution of $A^*$ and $S^*$, our non\hyphensym
  parametric measure of asymmetry and clumpiness (as defined in
  \S\ref{sec:modginim20}) measured for our AGN.  We fit each
  distribution with a Gaussian, and measure the centroids and FWHM of
  these distributions\longdash the distributions appear
  indistinguishable. The results of a K\hyphensym S test suggests that
  the two distributions are not independent.}
\label{fig:asymdist}
\end{figure}
\begingroup
\tiny
\begin{landscape}
\begin{figure}[h]
\begin{center}$
\begin{array}{cc}
\subfloat{
\includegraphics[width=2.0in]{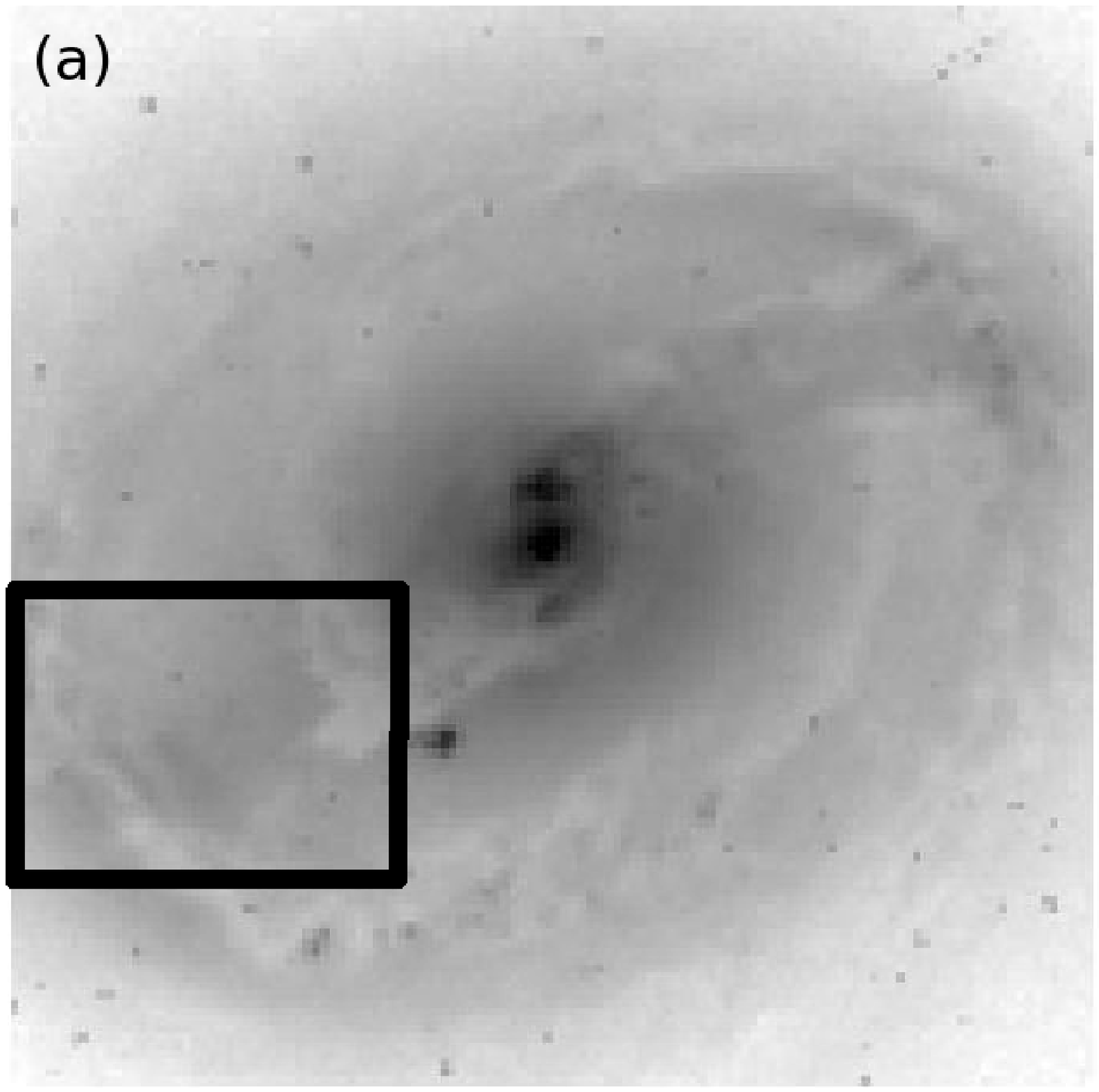}} &
\subfloat{\includegraphics[width=1.5in]{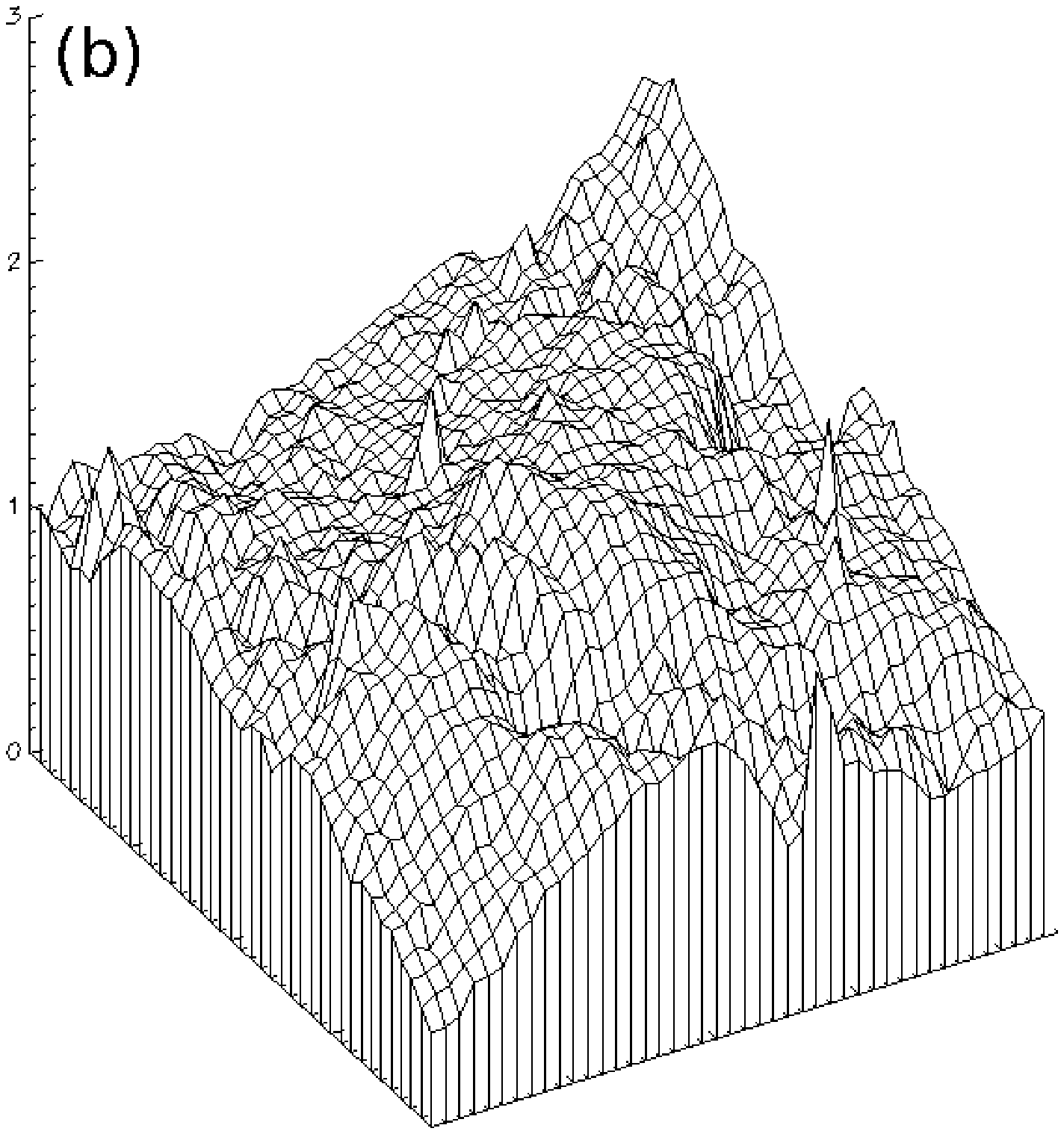}} \\
\subfloat{\framebox[2.5cm]{NGC 3081}} &
\subfloat{\framebox[5cm]{Spiral arm in NGC 3081}} \\
\subfloat{\includegraphics[width=1.5in]{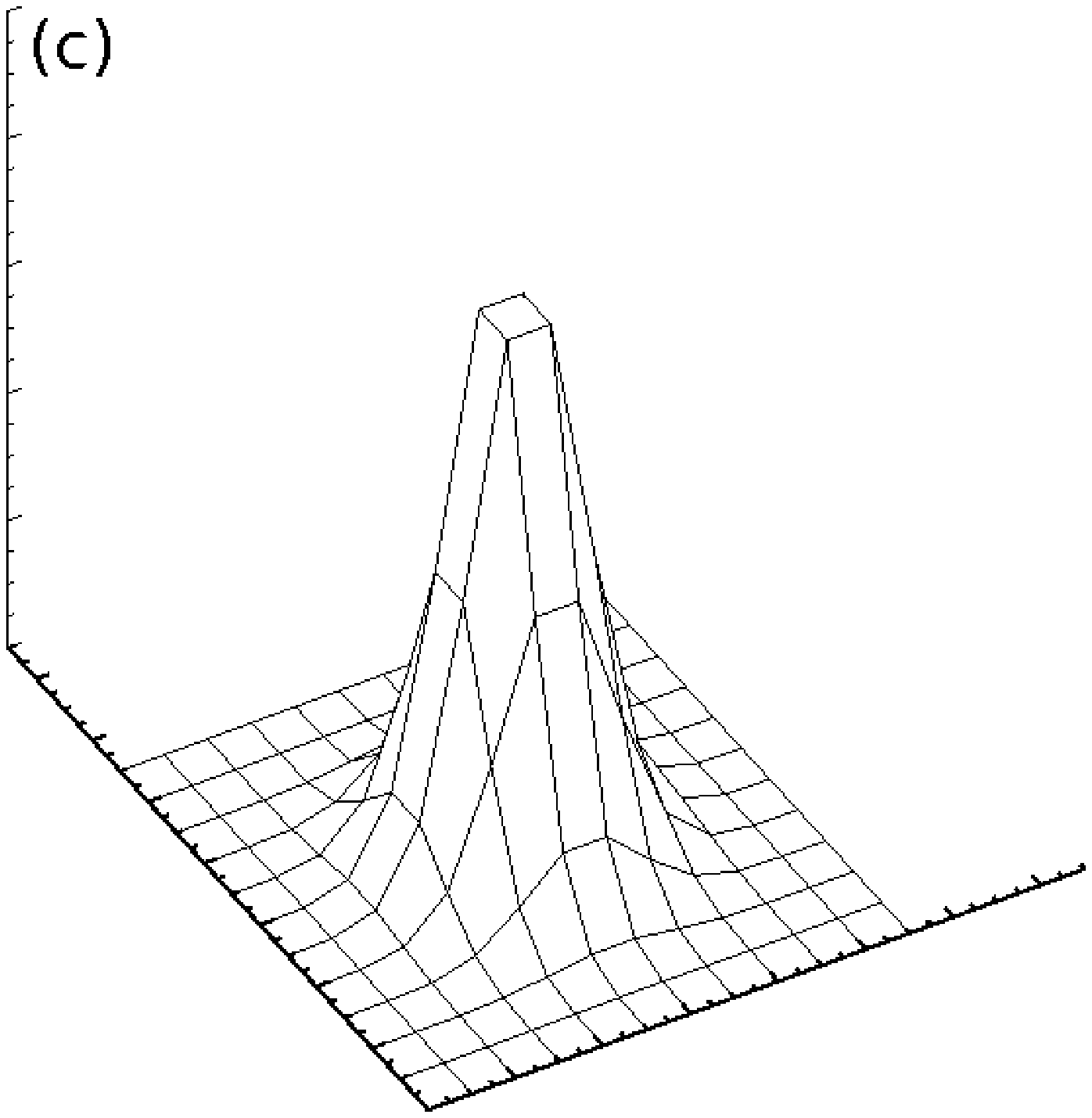}} &
\subfloat{\includegraphics[width=1.5in]{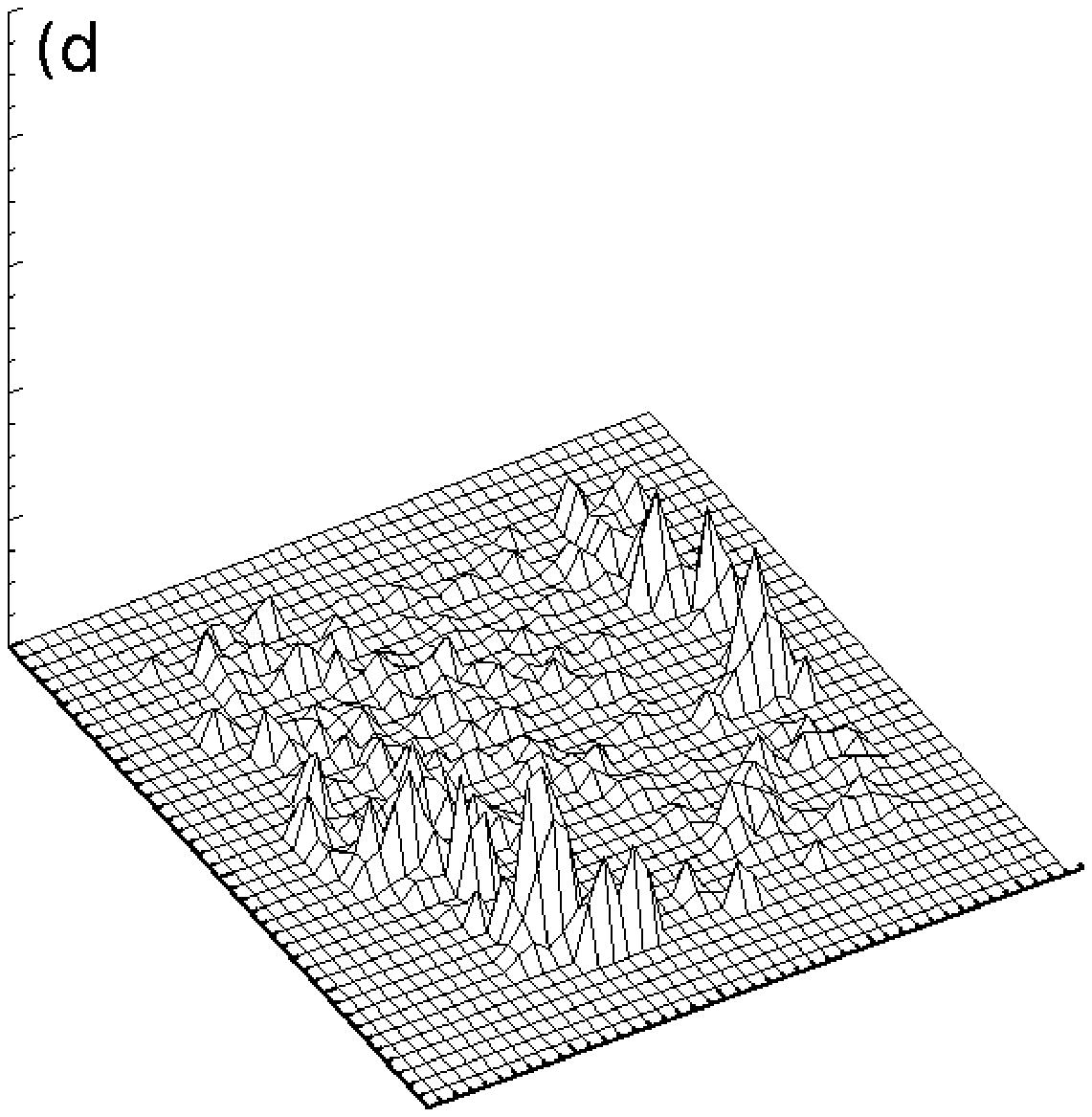}} \\
\subfloat{\framebox[4.cm]{Representative PSF}} &
\subfloat{\framebox[5.5cm]{Inverse Unsharp\hyphensym Mask Image}} \\ 
\end{array}$
\setlength{\abovecaptionskip}{+7pt}
\setlength{\belowcaptionskip}{+7pt}
\long\def\@makecaption#1#2{%
\vskip\abovecaptionskip
\sbox\@tempboxa{#1: #2}%
\ifdim \wd\@tempboxa >\hsize
#1: #2\par
\else
\global \@minipagefalse
\hb@xt@\hsize{\box\@tempboxa\hfil}%
\fi
\vskip\belowcaptionskip}
\makeatother
\caption{A cartoon representation of the {\it inverse
    unsharp\hyphensym mask} technique
  (see \S \ref{subsec:sextractmeth}) for detecting absorption of
  stellar light by dust and clumpy structures along the line\hyphensym
  of\hyphensym sight.  In panel (a), the 2kpc$\times$2kpc postage
  stamp of NGC3081 is provided; in this black and white image, black
  indicates relatively high signal and white indicates low signal. The
  thick black square in this figure emphasizes a spiral arm and
  inter-arm region with interesting dust features and morphology. A
  surface map ``zoom--in'' of this region is provided in panel (b);
  the arm is indicated by the ``trough'' extending in an arc from east
  to west in this image.  To produce the inverse unsharp\hyphensym
  mask image, we smoothed image (a) with a representative kernel
  (panel c) and divided the convolved image by the original image. In
  panel (d), we provide the unsharp\hyphensym mask surface map of the
  region in panel (b). It is apparent in panel (d) that the spiral arm
  region where dust absorption was most significant in panel (a) is
  now {\it sufficiently above the background} to be detectable
  using \sex~defined with an appropriate detection threshold.}
\end{center}
\label{fig:gaussconv}
\end{figure}
\end{landscape}
\endgroup
\setlength{\abovecaptionskip}{-7pt}
\setlength{\belowcaptionskip}{-7pt}
\long\def\@makecaption#1#2{%
\vskip\abovecaptionskip
\sbox\@tempboxa{#1: #2}%
\ifdim \wd\@tempboxa >\hsize
#1: #2\par
\else
\global \@minipagefalse
\hb@xt@\hsize{\box\@tempboxa\hfil}%
\fi
\vskip\belowcaptionskip}
\makeatother
\begin{figure}
\centering
\begin{tabular}{c}
\includegraphics[width=6in,angle=0,scale=1]{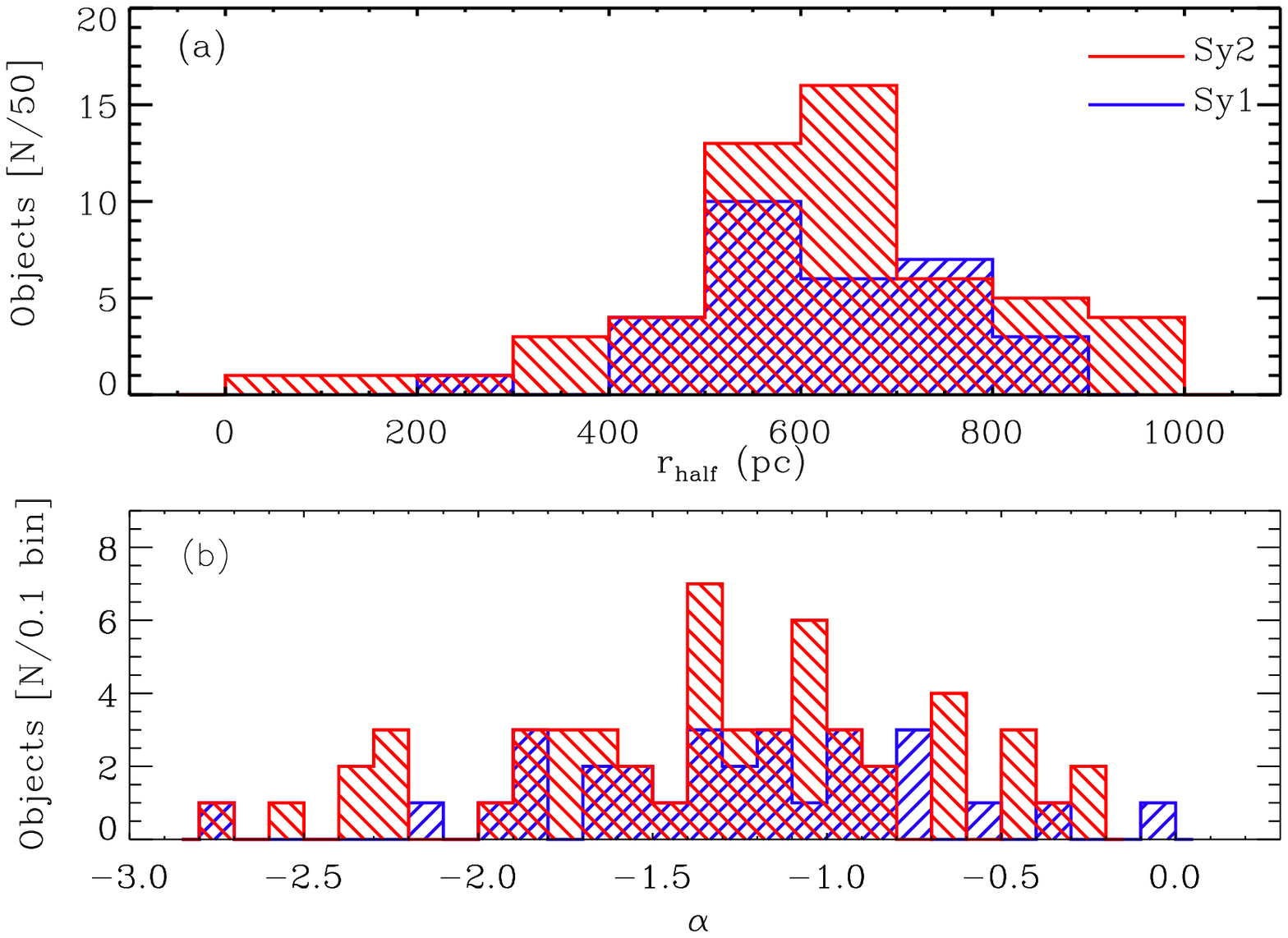} \\
\end{tabular}
\caption{The distribution of the best-fitting exponential slopes
  $\alpha$ to the object surface density profile and the half object
  radii of objects or features measured for objects in
  all \syall~galaxies detected by \sex~using the IUM technique
  (see \S\ref{subsec:sextractmeth}). We fit Gaussian functions to each
  of the distributions, and the results of a K\hyphensym S test
  confirms that the parents distribution from which the distributions
  were drawn are likely the same. This suggests that there is no
  significant difference between the azimuthally\hyphensym averaged
  spatial distribution of objects, and thus the distribution of dust
  features for the \syall~populations appears to be
  indistinguishable.}
\label{fig:alphadist}
\end{figure} 
\begin{figure}
\centering
\begin{tabular}{c}
\includegraphics[width=6in,angle=0,scale=1]{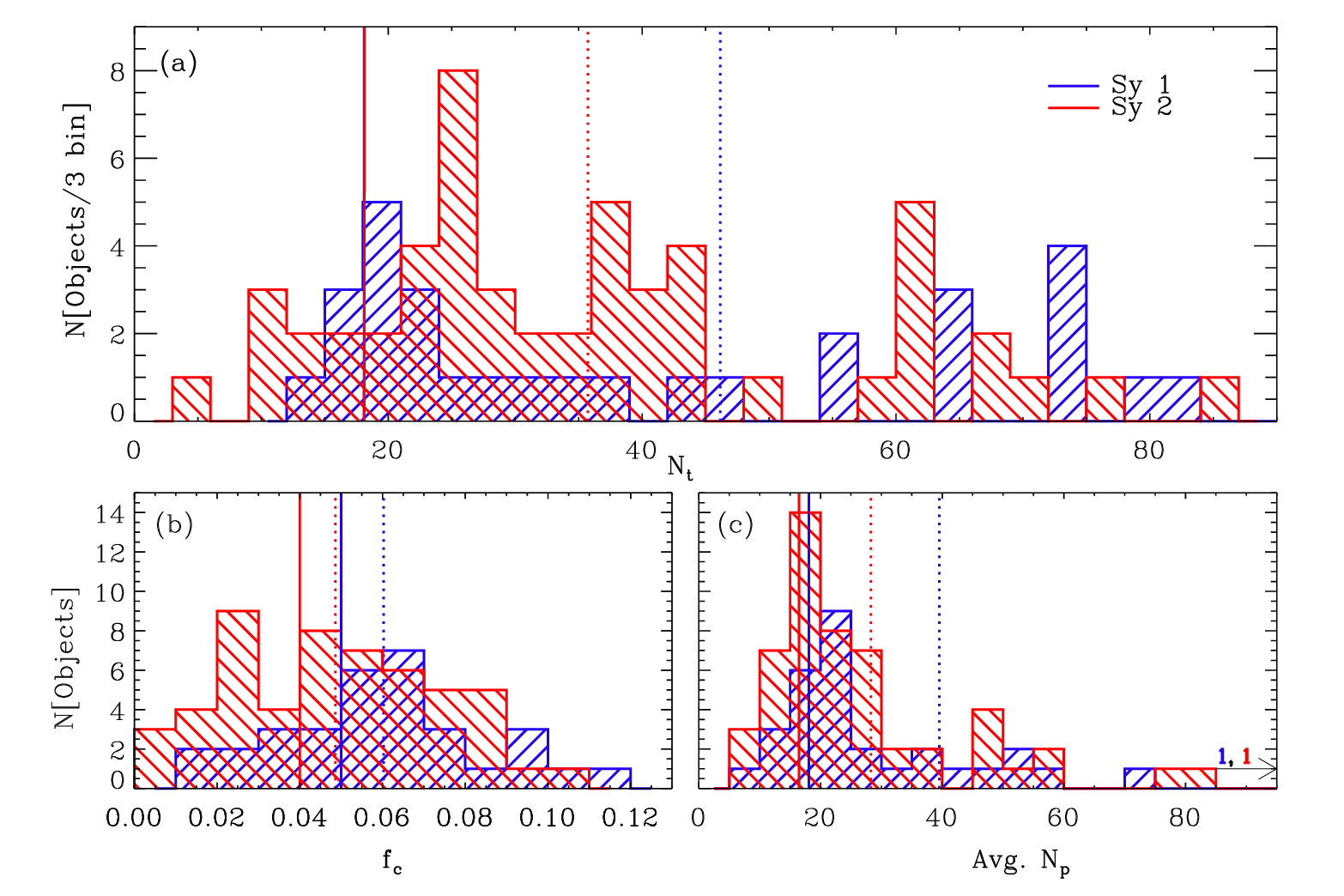} \\
\end{tabular}
\caption{The relative distributions of three statistics derived from
  the morphological technique discussed
  in \S \ref{subsec:sextractmeth}.  Panel (a)[Top]$\colon$ The number
  of morphological features $N_t$ detected in the core of each Seyfert
  galaxy. Panel (b)$\colon$ The distribution of covering fraction
  $f_c$ of dust features in the sample as defined
  in \S\ref{subsec:sexresults}.  $f_c$ is a measure of fraction of the
  total core image area that is associated with detected objects.  In
  general, \syall~host\hyphensym galaxies cover similar fractions of
  area of the host galaxy core. Panel (c)$\colon$ The distribution of
  the average number of pixels $N_p$ (i.e., object
  area$=N_p\times0\farcs1^2$; see \S\ref{subsec:sexresults}). Two
  galaxies (1 Sy1 and 1 Sy2) were detected with $N_p$\gsim\,arr90,
  indicated by the arrow.  In all panels, vertical dotted and solid
  lines indicate the mean and centroid (measured from the
  best\hyphensym fitting Gaussian, or Lorentzian in Panel a, function
  to each distribution) of the distributions.  The parameters of these
  fits, as well as the results of the results of the two-sample
  Kolmogorov-Smirnov tests of these distributions, are provided in
  Table \ref{tab:spartab2}.  Only for the distribution of total object
  number $N_t$ does the K\hyphensym S test suggest that the empirical
  distributions were not drawn from a common parent population.}
\label{fig:coverfrac}
\end{figure}

\clearpage
\newpage
\begingroup
\tiny
\begin{landscape}
\begin{longtable}{lccccccc}
\caption{AGN-Host Galaxy, Catalog}\\
\hline \hline
\multicolumn{1}{c}{ID$^a$} &
\multicolumn{1}{c}{Alt. ID$^b$} &
\multicolumn{1}{c}{R.A. (J2000)} & 
\multicolumn{1}{c}{Decl.(J2000)} & 
\multicolumn{1}{c}{Distance$^c$} &
\multicolumn{1}{c}{Sy. Type} &
\multicolumn{1}{c}{$\epsilon$} &
\multicolumn{1}{c}{NED Class} \\ \hline \hline
\endfirsthead

\multicolumn{8}{c}{AGN-Host Galaxy, (Continued)}\\
\hline \hline
\multicolumn{1}{c}{ID} &
\multicolumn{1}{c}{Alt. ID} &
\multicolumn{1}{c}{R.A.(J2000)} &
\multicolumn{1}{c}{Decl.(J2000)} & 
\multicolumn{1}{c}{Distance} &
\multicolumn{1}{c}{Sy. Type} &
\multicolumn{1}{c}{$\epsilon$} &
\multicolumn{1}{c}{NED Class} \\  \hline \hline
\endhead

\hline
\multicolumn{8}{l}{AGN-host galaxies (Continued)}\\
\endfoot
\hline \hline
\multicolumn{8}{l}{\textbf{Notes-} $^{\it a} \colon$ Object ID; $^{\it b} \colon$ NED preferred object ID; $^{\it c} \colon$ in Mpc using WMAP Year\hyphensym 7 cosmology \citep{K11}.}\\
\endlastfoot
ESO103-G035&IR1833-654&          18h38m20.3s&-65d25m39s            &             55.1&2.0& 0.63&S0?              \\
ESO137-G34&&		         16h35m14.1s&-58d04m48s                       &             37.8&2.0&            0.21&SAB0/a?(s)       \\  
ESO138-G1&&		         16h51m20.1s&-59d14m05s                        &             37.8&2.0&             0.50&E?               \\ 
ESO323-G77&&		         13h06m26.1s&-40d24m53s                       &             62.3&1.0&            0.33&(R)SAB0$^0$(rs)    \\ 
ESO362-G008&&		         05h11m09.0s&-34d23m35s                      &             51.5&2.0&           0.50&S0?              \\
ESO362-G018&&		         05h19m35.8s&-32d39m28s                      &             65.4&1.0&           0.33&SB0/a?(s) pec    \\
ESO373-G29&&		         09h47m43.5s&-32d50m15s                       &             38.6&2.0&            0.46&SB(rs)ab?        \\ 
FRL312&IC3639&                   12h40m52.9s&-36d45m21s                     &             45.2&2.0&          0.00&SB(rs)bc?        \\ 
FRL51&ESO140-G043&               18h44m54.0s&-62d21m53s                 &             58.8&1.0&      0.43&(R')SB(s)b?      \\ 
IR1249-131&NGC4748&              12h52m12.5s&-13d24m53s                &             65.2&1.0&     0.06& \nodata               \\ 
IR0450-032&PGC16226&             04h52m44.5s&-03d12m57s               &             60.7&2.0&  0.30& \nodata                \\ 
MARK352&&		         00h59m53.3s&+31d49m37s                          &             61.7&1.0&               0.50&SA0              \\
MARK1066&UGC02456&               02h59m58.6s&+36d49m14s                 &             49.8&2.0&      0.41&(R)SB0$^+$(s)      \\
MARK1126&NGC7450&                23h00m47.8s&-12d55m07s                  &             43.9&1.5&       0.00&(R)SB(r)a        \\
MARK1157&NGC0591&                01h33m31.3s&+35d40m06s                  &             63.0&2.0&       0.23&(R')SB0/a        \\ 
MARK1210&Phoenix&                08h04m05.9s&+05d06m50s                  &             56.0&1.0&       0.00&S?               \\
MARK1330&NGC4593&                12h39m39.4s&-05d20m39s                  &             37.2&1.0&       0.25&(R)SB(rs)b       \\ 
MARK270&NGC5283&                 13h41m05.8s&+67d40m20s                   &             43.0&2.0&        0.09&S0?              \\ 
MARK3&UGC03426&                  06h15m36.4s&+71d02m15s                    &             56.0&2.0&         0.11&S0?              \\
MARK313&NGC7465&                 23h02m01.0s&+15d57m53s                   &             27.0&2.0&        0.33&(R')SB0$^0$?(s)    \\ 
MARK348&NGC0262&                 00h48m47.1s&+31d57m25s                   &             62.4&2.0&        0.00&SA0/a?(s)        \\
MARK620&NGC2273&                 06h50m08.7s&+60d50m45s                   &             25.3&2.0&        0.21&SB(r)a?          \\ 
MARK686&NGC5695&                 14h37m22.1s&+36d34m04s                   &             58.5&2.0&        0.28&S?               \\
MARK744&NGC3786&                 11h39m42.6s&+31d54m33s                   &             36.9&1.8&        0.40&SAB(rs)a pec     \\ 
MARK766&NGC4253&                 12h18m26.5s&+29d48m46s                   &             53.6&1.5&        0.20&(R')SB(s)a?      \\
NGC1058&&		         02h43m30.0s&+37d20m29s                          &             7.10&2.0&               0.06&SA(rs)c          \\ 
NGC1068&MESSIER077&              02h42m40.7s&-00d00m48s                &             15.6&2.0&     0.15&(R)SA(rs)b       \\ 
NGC1125&&		         02h51m40.3s&-16d39m04s                          &             45.2&2.0&               0.50&(R')SB0/a?(r)    \\
NGC1241&&		         03h11m14.6s&-08d55m20s                          &             56.0&2.0&               0.39&SB(rs)b          \\ 
NGC1358&&		         03h33m39.7s&-05d05m22s                          &             55.7&2.0&               0.23&SAB0/a(r)        \\
NGC1365&&		         03h33m36.4s&-36d08m25s                          &             22.5&1.5&               0.44&SB(s)b           \\ 
NGC1386&&		         03h36m46.2s&-35d59m57s                          &             11.9&2.0&               0.61&SB0$^+$(s)         \\ 
NGC1566&&		         04h20m00.4s&-54d56m16s                          &             20.6&1.0&               0.20&SAB(s)bc         \\ 
NGC1667&&		         04h48m37.1s&-06d19m12s                          &             63.0&2.0&               0.22&SAB(r)c          \\
NGC1672&&		         04h45m42.5s&-59d14m50s                          &             18.2&2.0&               0.16& SB(r)bc\\ 
NGC2110&&		         05h52m11.4s&-07d27m22s                          &             32.1&2.0&               0.23&SAB0$^-$           \\
NGC2336&&		         07h27m04.1s&+80d10m41s                          &             30.3&2.0&               0.45&SAB(r)bc         \\ 
NGC2639&&		         08h43m38.1s&+50d12m20s                          &             46.0&1.9&               0.35&(R)SA(r)a?       \\
NGC2985&&		         09h50m22.2s&+72d16m43s                          &             18.1&1.9&               0.21&(R')SA(rs)ab     \\ 
NGC3081&&		         09h59m29.5s&-22d49m35s                          &             32.9&1.9&               0.23&(R)SAB0/a(r)     \\ 
NGC3185&&		         10h17m38.6s&+21d41m18s                          &             16.7&2.0&               0.49&(R)SB(r)a        \\ 
NGC3227&&		         10h23m30.6s&+19d51m54s                          &             15.8&1.5&               0.33&SAB(s)a pec      \\ 
NGC3393&&		         10h48m23.5s&-25d09m43s                          &             51.8&2.0&               0.09&(R')SB(rs)a?     \\
NGC3486&&		         11h00m23.9s&+28d58m30s                          &             9.34&2.0&               0.26&SAB(r)c          \\ 
NGC3516&&		         11h06m47.5s&+72d34m07s                          &             36.5&1.5&               0.23&(R)SB0$^0$?(s)     \\
NGC3608&&		         11h16m59.0s&+18d08m55s                          &             17.2&2.0&               0.18&E2               \\ 
NGC3718&&		         11h32m34.9s&+53d04m05s                          &             13.6&1.0&               0.50&SB(s)a pec       \\ 
NGC3783&&		         11h39m01.8s&-37d44m19s                          &             40.2&1.0&               0.10&(R')SB(r)ab      \\ 
NGC3982&&		         11h56m28.1s&+55d07m31s                          &             15.2&1.9&               0.11&SAB(r)b?         \\ 
NGC4051&&		         12h03m09.6s&+44d31m53s                          &             9.61&1.0&               0.25&SAB(rs)bc        \\ 
NGC4117&&		         12h07m46.1s&+43d07m35s                          &             12.8&2.0&               0.63&S0$^0$?            \\
NGC4303&MESSIER061&              12h21m54.9s&+04d28m25s                &             21.5&2.0&     0.10&SAB(rs)bc        \\ 
NGC4378&&		         12h25m18.1s&+04d55m31s                          &             35.2&2.0&               0.06&(R)SA(s)a        \\
NGC4395&&		         12h25m48.9s&+33d32m49s                          &             4.37&1.8&               0.16&SA(s)m?          \\ 
NGC4477&&		         12h30m02.2s&+13d38m12s                          &             18.6&2.0&               0.07&SB0(s)?          \\
NGC4507&&		         12h35m36.6s&-39d54m33s                          &             48.9&2.0&               0.23&(R')SAB(rs)b     \\
NGC4639&&		         12h42m52.4s&+13d15m27s                          &             13.9&1.0&               0.32&SAB(rs)bc        \\ 
NGC4698&&		         12h48m22.9s&+08d29m15s                          &             13.8&2.0&               0.37&SA(s)ab          \\ 
NGC4725&&		         12h50m26.6s&+25d30m03s                          &             16.5&2.0&               0.28&SAB(r)ab pec     \\ 
NGC4939&&		         13h04m14.4s&-10d20m23s                          &             42.9&2.0&               0.49&SA(s)bc          \\ 
NGC4941&&		         13h04m13.1s&-05d33m06s                          &             15.2&2.0&               0.47&(R)SAB(r)ab?     \\ 
NGC4968&&		         13h07m06.0s&-23d40m37s                          &             40.8&2.0&               0.52&(R')SAB0$^0$       \\
NGC5135&&		         13h25m44.1s&-29d50m01s                          &             56.8&2.0&               0.30&SB(s)ab          \\
NGC5273&&		         13h42m08.3s&+35d39m15s                          &             14.6&1.5&               0.06&SA0$^0$(s)         \\
NGC5347&&		         13h53m17.8s&+33d29m27s                          &             32.1&2.0&               0.23&(R')SB(rs)ab     \\ 
NGC5427&&		         14h03m26.1s&-06d01m51s                          &             36.1&2.0&               0.14&SA(s)c pec       \\ 
NGC5643&&		         14h32m40.7s&-44d10m28s                          &             16.4&2.0&               0.13&SAB(rs)c         \\ 
NGC5929&&		         15h26m06.2s&+41d40m14s                          &             34.3&2.0&               0.21&Sab? pec         \\ 
NGC5953&&		         15h34m32.4s&+15d11m38s                          &             27.0&2.0&               0.14&SAa? pec         \\ 
NGC6221&&		         16h52m46.1s&-59d13m07s                          &             20.6&1.0&               0.28&SB(s)c           \\ 
NGC6217&&		         16h32m39.2s&+78d11m53s                          &             18.7&2.0&               0.16&(R)SB(rs)bc      \\
NGC6300&&		         17h16m59.5s&-62d49m14s                          &             15.2&2.0&               0.33&SB(rs)b          \\ 
NGC6814&&		         19h42m40.6s&-10d19m25s                          &             21.4&1.5&               0.06&SAB(rs)bc        \\ 
NGC6890&&		         20h18m18.1s&-44d48m24s                          &             33.3&1.9&               0.20&SA(rs)b          \\ 
NGC6951&&		         20h37m14.1s&+66d06m20s                          &             19.5&2.0&               0.17&SAB(rs)bc        \\ 
NGC7213&&		         22h09m16.3s&-47d10m00s                          &             24.0&1.0&               0.09&SA(s)a?          \\ 
NGC7314&&		         22h35m46.2s&-26d03m02s                          &             19.6&1.9&               0.54&SAB(rs)bc        \\ 
NGC7410&&		         22h55m00.9s&-39d39m41s                          &             24.0&2.0&               0.69&SB(s)a           \\ 
NGC7469&&		         23h03m15.6s&+08d52m26s                          &             67.8&1.0&               0.26&(R')SAB(rs)a     \\
NGC7496&&		         23h09m47.3s&-43d25m41s                          &             22.6&2.0&               0.09&SB(s)b           \\ 
NGC7590&&		         23h18m54.8s&-42d14m21s                          &             21.6&2.0&               0.62&SA(rs)bc?        \\ 
NGC7682&&		         23h29m03.9s&+03d32m00s                          &             71.3&2.0&               0.08&SB(r)ab         \\
NGC7743&&		         23h44m21.1s&+09d56m03s                          &             23.5&2.0&               0.13&(R)SB0$^+$(s)      \\ 
NGC788&&		         02h01m06.4s&-06d48m56s                           &             56.4&1.0&               0.26&SA0/a?(s)        \\
TOL0109-383&NGC0424&             01h11m27.6s&-38d05m00s               &             48.7&2.0&               0.55&(R)SB0/a?(r)   
\label{tab:gallist}                                      
\end{longtable}
\end{landscape}
\endgroup

\begingroup
\tiny
\begin{landscape}
\begin{longtable}{lccccc}
\caption{AGN-Host Galaxy, Catalog}\\
\hline \hline
\multicolumn{1}{l}{ID$^a$} &
\multicolumn{1}{c}{Morpho.}&
\multicolumn{1}{c}{Ancillary} &
\multicolumn{1}{c}{Morpho.} &
\multicolumn{1}{c}{$\alpha$ ($\chi^2$)} &
\multicolumn{1}{c}{Half Object} \\
\multicolumn{1}{l}{} &
\multicolumn{1}{c}{Class A} &
\multicolumn{1}{c}{Class.} &
\multicolumn{1}{c}{Class B} &
\multicolumn{1}{c}{} &
\multicolumn{1}{c}{Radius} \\ \hline \hline
\endfirsthead
\multicolumn{6}{c}{AGN-Host Galaxy, (Continued)}\\
\hline \hline
\multicolumn{1}{l}{ID$^a$} &
\multicolumn{1}{c}{Morpho.}&
\multicolumn{1}{c}{Ancillary} &
\multicolumn{1}{c}{Morpho.} &
\multicolumn{1}{c}{$\alpha$ ($\chi^2$)} &
\multicolumn{1}{c}{Half Object}\\
\multicolumn{1}{l}{} &
\multicolumn{1}{c}{Class A} &
\multicolumn{1}{c}{Class.} &
\multicolumn{1}{c}{Class B} &
\multicolumn{1}{c}{} &
\multicolumn{1}{c}{Radius} \\ \hline \hline
\endhead

\hline
\multicolumn{6}{l}{AGN-host galaxies (Continued)}\\
\endfoot
\hline \hline
\multicolumn{6}{l}{\textbf{Notes-} Combined results from qualitative and quantitative analyses are provided.}\\
\multicolumn{6}{l}{Cols.2-3$\colon$Morphological classifiers were adopted from \cite{M98}}\\
\multicolumn{6}{l}{\,\,\,\,\,\,and are defined as follows$\colon$ {\sc Class A}--F/W=Filaments/Wisps; DI=Irregular Dust;}\\
\multicolumn{6}{l}{\,\,\,\,\,\,DC= dust lane passing close to, or bisecting, center; D-[direction]=dust lanes}\\
\multicolumn{6}{l}{\,\,\,\,\,on one side of major axis. {\sc Ancillary}--B=bar; CL=cluster, lumpy H II region, knots;}\\
\multicolumn{6}{l}{\,\,\,\,\,\,E/S0=Elliptical; R=ring. We do not use the ``Normal'' classifier.  For details, \ref{sec:visclass}.}\\
\multicolumn{6}{l}{Col 4.$\colon$ {\sc Class B}--Classifiers defined to specifically characterize the}\\
\multicolumn{6}{l}{\,\,\,\,\,\,dust morphology in the cores of galaxies$\colon$}\\
\multicolumn{6}{l}{\,\,\,\,\,\,{\it 1}$\colon$``Spiral''; Absorption''; {\it 2}$\colon$``Bar''; and }\\
\multicolumn{6}{l}{\,\,\,\,\,\,{\it 3}$\colon$ ``Dust Specific Notes'' -- {\it s} \mbox{or} {\it i}$\colon$``spiral'' or ``irregular'' \& {\it m} \mbox{or} {\it l}$\colon$ ``High''}\\
\multicolumn{6}{l}{\,\,\,\,\,\, or ``Low Extinction''. The full description of the classifiers is provided in \S \ref{sec:visclass}}\\
\multicolumn{6}{l}{Col. 5$\colon$ Slope ($\alpha$) of best\hyphensym fit line to the object surface density profile defined in \S \ref{subsec:sexresults}}\\
\multicolumn{6}{l}{\,\,\,\,\,\,Reduced $\chi^2$ is provided in parantheses.}\\
\multicolumn{6}{l}{Col. 6$\colon$Half-Object Radius (pc) with uncertainty.  Here,``\nodata'' indicates that a galaxy}\\
\multicolumn{6}{l}{\,\,\,\,\,\,had an insufficiently few objects to measure the radius accurately.}\\
\endlastfoot

IR1833-654&DI,F/W&B&2,3il&$-$0.45(0.21)&967.53$\pm$54.17\\
ESO137-G34&DI,DC&CL&3im&$-$1.51(0.23)&927.31$\pm$37.47\\
ESO138-G1&D-NW&-&3im&$-$1.16(0.17)&612.37$\pm$37.45\\
ESO323-G77&F/W&R,CL&1,3s&$-$1.90(0.06)&350.03$\pm$61.09\\
ESO362-G18&D-SW,DC,F/W&R&3sm&$-$1.04(0.16)&789.24$\pm$64.02\\
ESO362-G8&DC,DI&-&3il&$-$0.40(0.00)&\nodata\\
ESO373-G29&DI,D-NE,F/W&B&3il&$-$0.69(0.22)&823.03$\pm$38.28\\
FRL312&DC,F/W&B&2,3im&$-$0.94(0.31)&562.27$\pm$44.66\\
FRL51&DI&CL&3il&$-$2.14(0.07)&392.60$\pm$57.75\\
IR1249-131&DC,F/W&R,CL&3sl&$-$0.93(0.06)&589.14$\pm$59.56\\
IR0450-032&DC&R&3im&$-$2.52(0.49)&604.99$\pm$63.79\\
MARK352&-&E/S0&-&$-$1.45(0.28)&296.90$\pm$60.49\\
MARK1066&F/W&B,CL&2,3sm&$-$2.25(0.19)&688.47$\pm$49.10\\
MARK1126&DI,F/W&B,R&2,3il&$-$1.14(0.32)&821.58$\pm$43.45\\
MARK1157&DI,F/W,D-NE&B,R&2,3sl&$-$1.09(0.00)&533.15$\pm$61.70\\
MARK1210&F/W&CL,R&3sl&$-$0.81(0.67)&919.57$\pm$55.01\\
MARK1330&F/W,DC&R&3sm&$-$1.53(0.24)&879.67$\pm$36.88\\
MARK270&F/W,D-S,DC&B&3sl&$-$1.38(0.19)&474.55$\pm$42.56\\
MARK3&DI,D-NE&B,CL&2,3im&$-$1.36(0.26)&817.55$\pm$55.07\\
MARK313&DI,DC,F/W&B,CL&3im&$-$1.54(0.73)&935.46$\pm$26.98\\
MARK348&F/W&-&3sl&$-$1.68(0.38)&646.81$\pm$61.17\\
MARK620&F/W,DI,D-N,&B,R,CL&3sm&$-$1.46(0.35)&551.32$\pm$25.24\\
MARK686&D-W,DC,F/W&-&3sm&$-$2.40(0.42)&485.03$\pm$57.41\\
MARK744&DI,DC&R,CL&3sl&$-$1.36(0.33)&528.30$\pm$36.61\\
MARK766&DI&-&3il&$-$1.97(0.54)&727.96$\pm$52.74\\
NGC1058&F/W,DI&CL&1,3s&$-$0.90(1.28)&699.35$\pm$7.144\\
NGC1068&F/W,DI&CL&1,3sm&$-$0.22(0.85)&855.87$\pm$15.64\\
NGC1125&DC,D-SW,DI&-&3im&$-$1.36(0.19)&592.58$\pm$44.69\\
NGC1241&DC,F/W&CL,R&3sm&$-$1.72(0.00)&535.62$\pm$55.09\\
NGC1358&DC,DI&-&3il&$-$1.62(0.24)&594.62$\pm$54.77\\
NGC1365&DC,F/W&CL&3im&$-$1.23(0.94)&620.82$\pm$22.46\\
NGC1386&D-NW,F/W,DC&-&1,3sm&$-$1.08(0.39)&543.61$\pm$11.95\\
NGC1566&F/W,DC&R&3sm&$-$1.00(0.22)&715.97$\pm$20.66\\
NGC1667&F/W,DC&-&3sl&$-$2.34(0.46)&641.56$\pm$61.70\\
NGC1672&F/W,DC&CL&1,3sm&$-$1.68(2.35)&558.84$\pm$18.29\\
NGC2110&F/W,DC,DI,D-N&-&1,3sm&$-$0.96(0.12)&566.33$\pm$31.96\\
NGC2336&DI&E/S0&3il&$-$0.83(0.31)&794.94$\pm$30.19\\
NGC2639&F/W,DC,D-NE&B&3im&$-$1.34(0.02)&513.96$\pm$45.49\\
NGC2985&F/W,DC&-&3s&$-$0.37(0.91)&718.04$\pm$18.17\\
NGC3081&DI,F/W&B,R,CL&1,2,3sm&$-$1.20(0.49)&738.19$\pm$32.72\\
NGC3185&DC,DI&R&3i&$-$1.99(0.08)&132.99$\pm$16.73\\
NGC3227&DI,DC&-&3il&$-$0.87(0.27)&549.46$\pm$15.91\\
NGC3393&F/W,DI,DC&B,CL&2,3sm&$-$2.27(0.04)&472.74$\pm$51.05\\
NGC3486&F/W,DI&R,CL&1,3sl&$-$0.65(2.00)&764.01$\pm$9.387\\
NGC3516&DI&-&3il&$-$1.55(0.04)&474.70$\pm$36.22\\
NGC3608&-&E/S0&-&$-$0.70(0.30)&851.63$\pm$17.23\\
NGC3718&DI,DC,D-SW&-&3im&$-$1.37(0.82)&492.53$\pm$13.66\\
NGC3783&DI&E/S0&3il&$-$1.70(0.21)&610.91$\pm$39.84\\
NGC3982&F/W,DI&R,CL&1,3sm&$-$0.02(0.69)&508.14$\pm$15.25\\
NGC4051&DC,DI&-&3im&$-$0.96(0.64)&800.02$\pm$9.650\\
NGC4117&DI,F/W,DC&R&3im&$-$0.25(1.14)&683.86$\pm$12.85\\
NGC4303&F/W&R&1,3sm&$-$1.77(1.28)&391.61$\pm$21.50\\
NGC4378&DI&-&3i&$-$1.70(0.17)&465.32$\pm$35.00\\
NGC4395&-&CL&-&$-$1.36(5.03)&526.02$\pm$4.402\\
NGC4477&DC,D-E&-&3il&$-$1.24(0.21)&540.95$\pm$18.62\\
NGC4507&D-S,DI&-&3im&$-$1.80(0.39)&592.75$\pm$48.20\\
NGC4639&F/W&B&2,3sl&$-$0.71(0.58)&793.61$\pm$14.00\\
NGC4698&DI&E/S0&3i&$-$1.02(0.48)&664.53$\pm$13.88\\
NGC4725&DI&E/S0&3il&$-$0.93(1.41)&743.22$\pm$16.58\\
NGC4939&D-W,F/W&B&2,3im&$-$1.15(0.16)&697.48$\pm$42.44\\
NGC4941&D-E,DI&-&3il&$-$0.49(1.08)&863.44$\pm$15.24\\
NGC4968&DC,F/W,D-NE&-&3im&$-$1.35(0.09)&609.44$\pm$40.38\\
NGC5135&DI&R,CL&3sm&$-$2.72(0.38)&578.57$\pm$55.80\\
NGC5273&F/W,DC,DI&-&3il&$-$1.87(0.79)&212.56$\pm$14.64\\
NGC5347&F/W,DI&R&3im&$-$0.44(0.35)&792.55$\pm$31.96\\
NGC5427&F/W&R,CL&1,3sm&$-$1.83(0.26)&789.46$\pm$35.80\\
NGC5643&F/W&CL&1,3sl&$-$1.18(1.25)&663.69$\pm$16.48\\
NGC5929&DC,DI&-&3il&$-$1.06(0.09)&667.89$\pm$34.09\\
NGC5953&F/W&CL&1,3sm&$-$1.32(1.63)&667.67$\pm$26.94\\
NGC6217&DI,DC&CL&3sm&$-$1.09(0.16)&694.41$\pm$20.58\\
NGC6221&DI,DC,D-SE&CL&3il&$-$0.78(0.21)&624.49$\pm$18.71\\
NGC6300&DI,D-SW&CL&3im&$-$1.08(0.20)&714.21$\pm$15.25\\
NGC6814&F/W,DC&B&2,3sl&$-$0.74(0.25)&585.77$\pm$21.46\\
NGC6890&F/W&CL&1,3sm&$-$1.11(0.37)&767.87$\pm$33.10\\
NGC6951&DI,F/W&R,B,CL&2,3sm&$-$1.38(0.67)&545.79$\pm$19.56\\
NGC7213&F/W&-&3sl&$-$1.22(0.26)&620.10$\pm$24.02\\
NGC7314&D-E,DC,F/W&-&3im&$-$0.58(0.29)&580.60$\pm$19.62\\
NGC7410&DC,F/W,D-NW&-&3im&$-$1.81(0.07)&450.93$\pm$24.03\\
NGC7469&F/W&R,B,CL&3sm&$-$2.75(0.07)&519.75$\pm$66.29\\
NGC7496&DC,DI,D-NW&CL&3im&$-$1.28(0.36)&641.13$\pm$22.63\\
NGC7590&F/W,D-NW&CL&3im&$-$1.28(0.27)&684.94$\pm$21.63\\
NGC7682&DI&CL&3i&$-$2.25(0.32)&691.53$\pm$69.57\\
NGC7743&F/W&-&3il&$-$0.68(0.34)&557.79$\pm$23.47\\
NGC788&F/W,D-S&-&3sl&$-$1.86(0.35)&415.99$\pm$55.44\\
TOL0109-383&D-SE,F/W&-&3im&$-$1.87(0.46)&668.87$\pm$48.05
\label{tab:fitpars}
\end{longtable}
\end{landscape}
\endgroup
\begingroup
\tiny
\begin{longtable}{c|cccccccc}
\caption{Morphology Comparison I$\colon$~$C^*A^*S^*$ \& $G^*$\hyphensym$M^*_{20}$ Technique}\\
\hline \hline
\multicolumn{1}{c}{Morpho.} &
\multicolumn{2}{c}{Sy1}  & 
\multicolumn{1}{c}{} &
\multicolumn{2}{c}{Sy2}  & 
\multicolumn{1}{c}{} &
\multicolumn{2}{c}{K-S Test}  \\ \cline{2-3} \cline{5-6} \cline{8-9}
\multicolumn{1}{c}{Param.} &
\multicolumn{1}{c}{Centroid} &
\multicolumn{1}{c}{FWHM} &
\multicolumn{1}{c}{} &
\multicolumn{1}{c}{Centroid} &
\multicolumn{1}{c}{FWHM} & 
\multicolumn{1}{c}{} &
\multicolumn{1}{c}{{\it d}} & 
\multicolumn{1}{c}{{\it p}} \\ \hline \hline 
\endhead

\hline \hline
\multicolumn{9}{l}{\textbf{Notes-} The parameters of the best\hyphensym fit Gaussian function to the distribution}\\
\multicolumn{9}{l}{\,\,\,\,\,\,\,\,\,\,\,of morphological parameters measured for\syall, respectively, are}\\
\multicolumn{9}{l}{\,\,\,\,\,\,\,\,\,\,\,provided here. For more details and a discussion of the implications of}\\
\multicolumn{9}{l}{\,\,\,\,\,\,\,\,\,\,\,these results in our test of the Unified Model, see \S\ref{sec:modginim20}.}\\
\endfoot
$G^*$       &0.39 & 0.26 & & 0.42 & 0.35 & & 0.28 & 0.09 \\
$M^*_{20}$   & $-$1.89 & 0.74 & & $-$2.01 & 1.24 & & 0.29 & 0.08 \\
$C^*$       & 3.15  & 1.27 & & 3.57 & 2.01 & & 0.38 & 0.01 \\
$A^*$       & 0.08 & 0.08 & & 0.12 & 0.09 & & 0.09 & 0.99 \\
$S^*$       & 0.08 & 0.13 & &  0.066 & 0.13 & & 0.19 & 0.91 
\label{tab:spartab1}
\end{longtable}

\endgroup
\begingroup
\tiny
\begin{longtable}{c|cccccccc}
\caption{Morphology Comparison II$\colon$\sex~Technique}\\
\hline \hline
\multicolumn{1}{c}{Morpho.} &
\multicolumn{2}{c}{Sy1}  & 
\multicolumn{1}{c}{} &
\multicolumn{2}{c}{Sy2}  & 
\multicolumn{1}{c}{} &
\multicolumn{2}{c}{K-S Test}  \\ \cline{2-3} \cline{5-6} \cline{8-9}
\multicolumn{1}{c}{Param.} &
\multicolumn{1}{c}{Centroid} &
\multicolumn{1}{c}{FWHM} &
\multicolumn{1}{c}{} &
\multicolumn{1}{c}{Centroid} &
\multicolumn{1}{c}{FWHM} & 
\multicolumn{1}{c}{} &
\multicolumn{1}{c}{{\it d}} & 
\multicolumn{1}{c}{{\it p}} \\ \hline \hline
\endhead

\hline \hline
\multicolumn{9}{l}{\textbf{Notes-} The parameters of the best\hyphensym fit Gaussian function to the distribution}\\
\multicolumn{9}{l}{\,\,\,\,\,\,\,\,\,\,\,of morphological parameters measured for\syall, respectively, are}\\
\multicolumn{9}{l}{\,\,\,\,\,\,\,\,\,\,\,provided here.  The HWHM value is provided for $N_t$ because  a Lorentzian,}\\
\multicolumn{9}{l}{\,\,\,\,\,\,\,\,\,\,\,Lorentzian, not a Gaussian, function was fit to the measured distribution}\\
\multicolumn{9}{l}{\,\,\,\,\,\,\,\,\,\,\,of this parameter. For more details and a discussion of the}\\
\multicolumn{9}{l}{\,\,\,\,\,\,\,\,\,\,\,implication of these results in our test of the Unified Model, see \S\ref{subsec:sextractmeth}.}\\

\endfoot
$\alpha$     &$-$1.27 & 0.79 & & $-1.30$ & 0.88 & & 0.14 & 0.69  \\
$r_{half}$       & 570  & 226 & & 572 & 195 & & 0.33 & 0.43 \\
$N_t$       & 18 & 4 & & 18 & 15 & & 0.35 & 0.01 \\
$f_c$       & 0.05 & 0.04 & & 0.04 & 0.05 & & 0.37 & 0.16 \\
$N_p$       & 18 & 7 & & 16 & 11 & & 0.12 & 0.98 
\label{tab:spartab2}
\end{longtable}
\endgroup
\setlength{\abovecaptionskip}{0pt}
\setlength{\belowcaptionskip}{0pt}
      
\long\def\@makecaption#1#2{%
\vskip\abovecaptionskip
\sbox\@tempboxa{#1: #2}%
\ifdim \wd\@tempboxa >\hsize
#1: #2\par
\else
\global \@minipagefalse
\hb@xt@\hsize{\box\@tempboxa\hfil}%
\fi
\vskip\belowcaptionskip}
\makeatother
 
\begin{figure}
\centering
\begin{tabular}{c}
\includegraphics[width=6in,angle=0,scale=0.55]{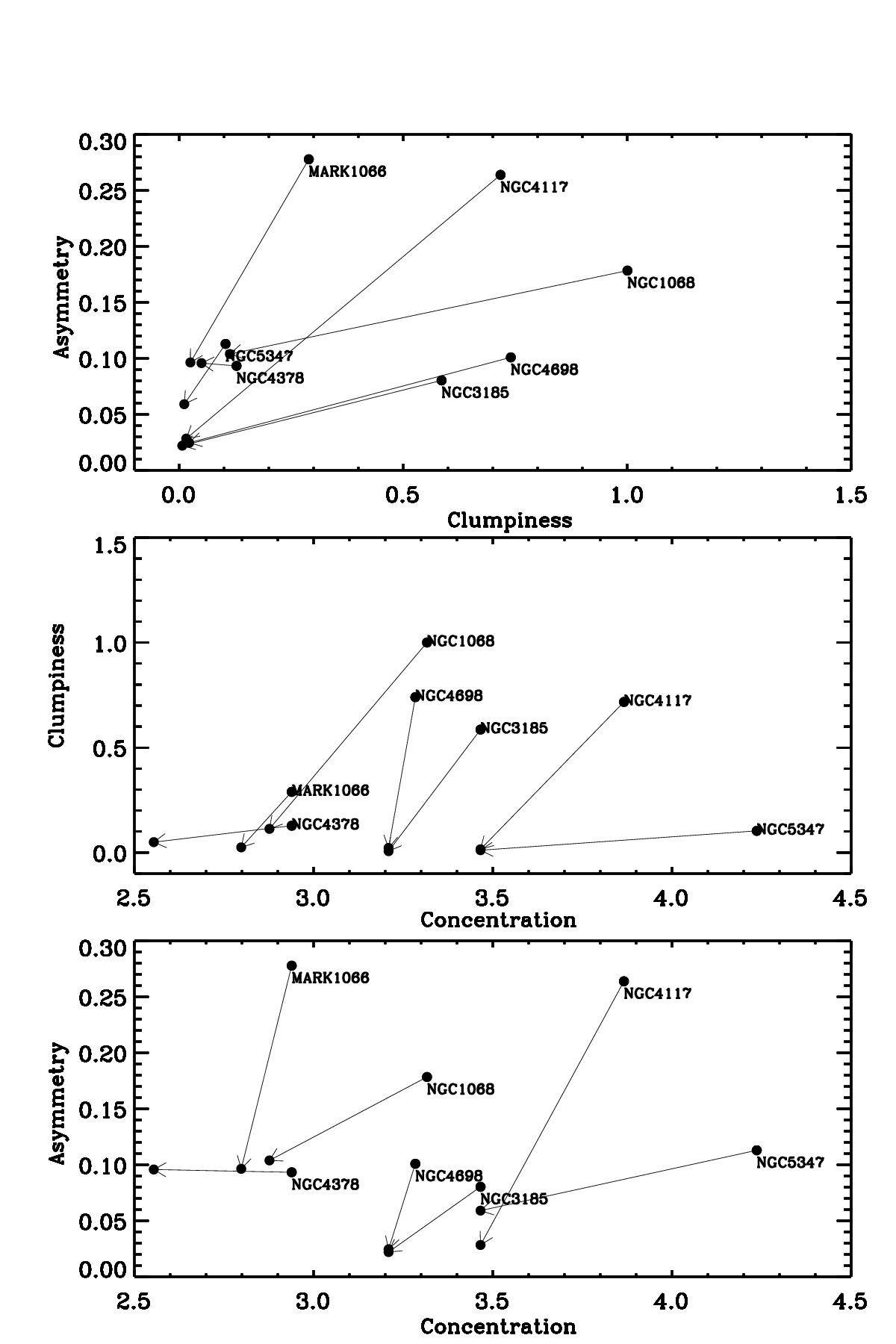} \\
\end{tabular}
\caption{$C^*, A^*,$ and $S^*$ measured for the core (r$<$1kpc)of 7 AGN
  using {\it HST} F606W and SDSS r$^{\prime}$ images. Line segments connect
  the measured values for each galaxy, and the vector points away from
  the parameter value measured from the {\it HST} image.}
\label{fig:appendixa}
\end{figure} 
\begin{deluxetable}{cc}
\tabletypesize{\scriptsize} \tablecaption{Distance Dependence$^a$}\tablenotetext{a}{\longdash$\delta$ quantifies the dispersion of the parameter at native and artificially redshift spatial resolution.  See Appendix B for details.}

\tablewidth{0pt} 
\tablehead{\colhead{Parameter} & \colhead{$\delta$}}
\startdata 
$G^*$ & 1.5\% \\
$M^*_{20}$ & 0.8\% \\
$C^*$ & 10.9\% \\
$A^*$ & 2.2\% \\
$S^*$ & 14.1\%  \\
\enddata
\label{tab:appendixb}
\end{deluxetable}
    
\begingroup
\tiny
\begin{figure}[h]
\begin{center}$
\begin{array}{cc}
\subfloat{\label{sfig:2a}\includegraphics[width=3in]{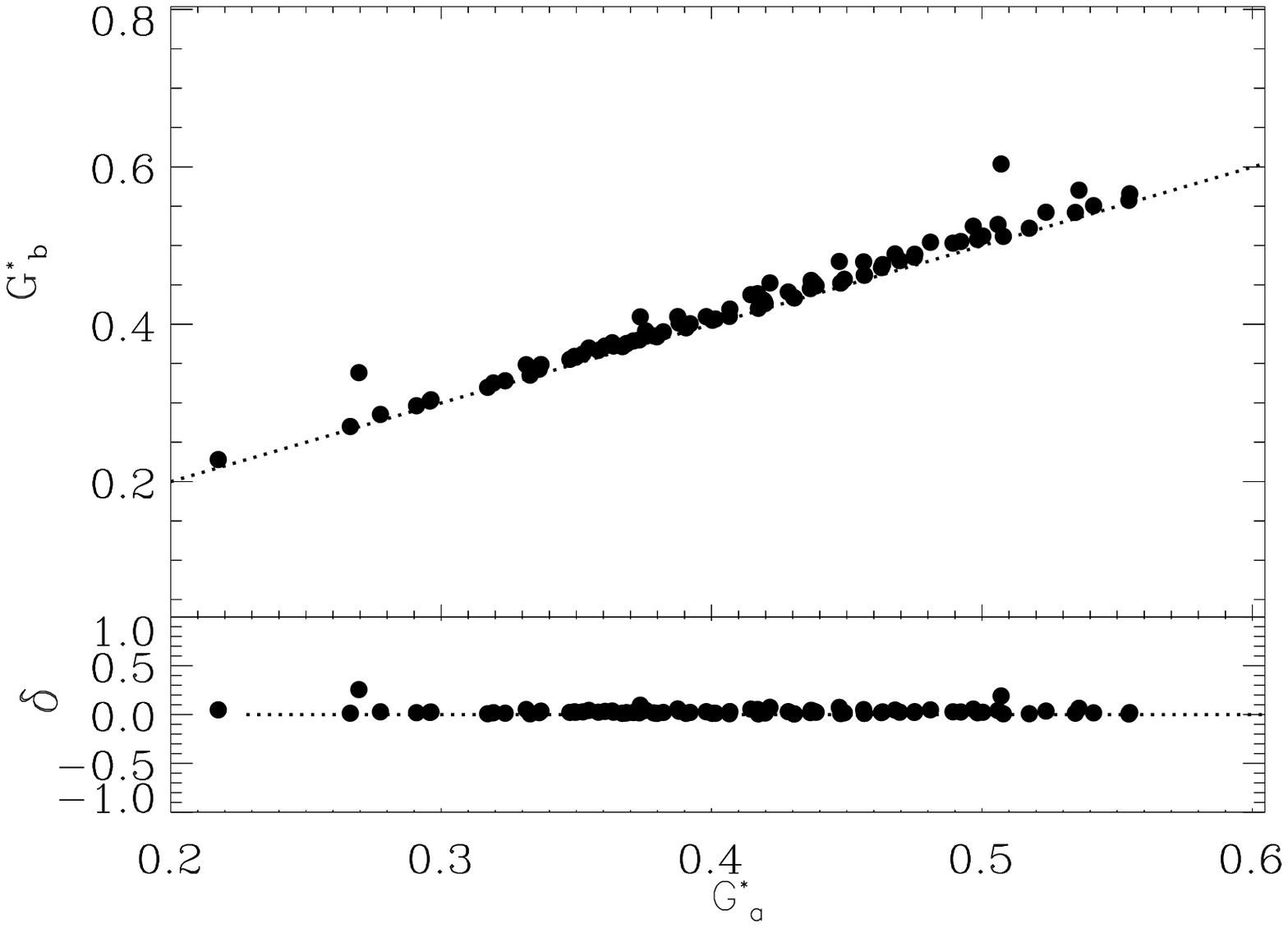}} &
\subfloat{\label{sfig:1a}\includegraphics[width=3in]{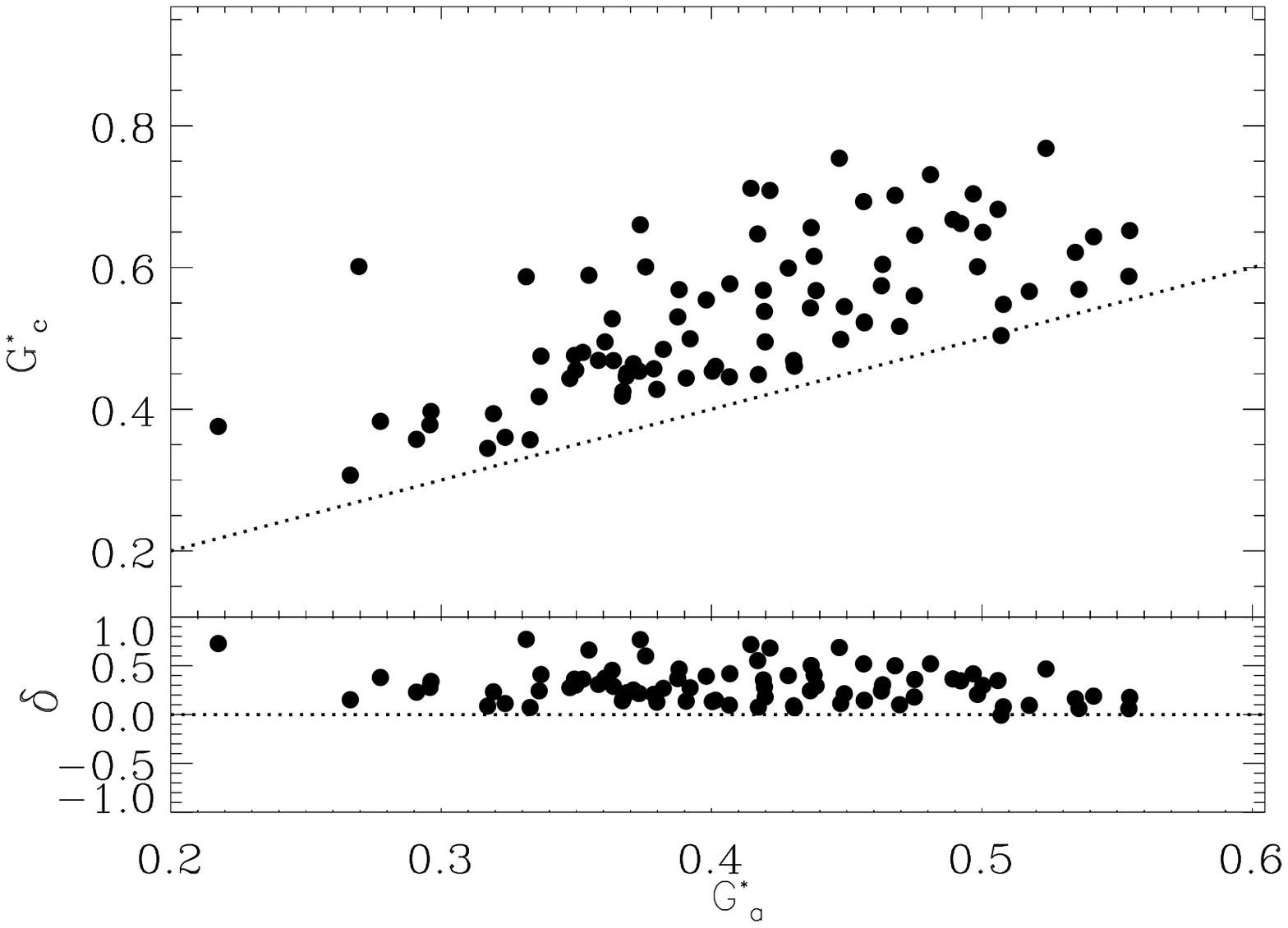}} \\
\end{array}$

\caption{We discuss in Appendix C the robustness of the parameters to the estimate of sky background. $G^*$ was measured for the galaxies in images produced for three assumptions of the zodiacal background surface brightness equal to $\colon$ (1) zero, $G^*_a$; (2) estimated from Windhorst et al. (in prep.), $G^*_b$, and (3) a (hypothetical) 10$\times$ {\it larger} than Windhorst et al.,$G^*_c$.  In the left (right) panel, we show the measured dispersion ($\delta=\frac{G^*_x-G^*_a}{G^*_a}$), where {\it X} indicates measurements in scenarios (2) and (3).  We measure a significant difference ($>20\%$) only for scenario (3). See Appendix C for more details.}
\end{center}
 \label{fig:appendixc}
\addtocounter{figure}{1}
\end{figure}
\endgroup

\end{document}